 \documentclass{emulateapj}                                                    

\usepackage{amsmath, latexsym, color, verbatim}
\usepackage{graphicx,epsfig,epstopdf,natbib}

\slugcomment{DRAFT: \today}

\begin{document}

\newcommand{\htwo}{\textrm{H\,{\sc ii}}}
\newcommand{\Ha}{\textrm{H\,{$\alpha$}}}
\newcommand{\otwo}{[\ion{O}{2}]}

\shorttitle{MaNGA Observing Strategy}
\shortauthors{Law et al.}

\title{Observing Strategy for the SDSS-IV/M{\rm {\footnotesize a}}NGA IFU Galaxy Survey}

\author{David R.~Law\altaffilmark{1}, 
Renbin Yan\altaffilmark{2},
Matthew A.~Bershady\altaffilmark{3},
Kevin Bundy\altaffilmark{4},
Brian Cherinka\altaffilmark{5},
Niv Drory\altaffilmark{6},
Nicholas MacDonald\altaffilmark{7},
Jos\'e R.~S\'anchez-Gallego\altaffilmark{2},
David A.~Wake\altaffilmark{3,8}
Anne-Marie Weijmans\altaffilmark{9},
Michael R.~Blanton\altaffilmark{10}, 
Mark A.~Klaene\altaffilmark{11},
Sean M.~Moran\altaffilmark{12},
Sebastian F.~Sanchez\altaffilmark{13},
Kai~Zhang\altaffilmark{2}
}

\altaffiltext{1}{Space Telescope Science Institute, 3700 San Martin Drive, Baltimore, MD 21218, USA (dlaw@stsci.edu)}
\altaffiltext{2}{Department of Physics and Astronomy, University of Kentucky, 505 Rose Street, Lexington, KY 40506-0055, USA}
\altaffiltext{3}{Department of Astronomy, University of Wisconsin-Madison, 475 N. Charter Street, Madison, WI, 53706, USA}
\altaffiltext{4}{Kavli Institute for the Physics and Mathematics of the Universe, Todai Institutes for Advanced Study, the University of Tokyo, Kashiwa, Japan 277-8583 (Kavli IPMU, WPI)}
\altaffiltext{5}{Dunlap Institute for Astronomy and Astrophysics, University of Toronto, 50 St. George Street, Toronto, Ontario M5S 3H4, Canada}
\altaffiltext{6}{McDonald Observatory, Department of Astronomy, University of Texas at Austin, 1 University Station, Austin, TX 78712-0259, USA}
\altaffiltext{7}{Department of Astronomy, Box 351580, University of Washington, Seattle, WA 98195, USA}
\altaffiltext{8}{Department of Physical Sciences, The Open University, Milton Keynes, MK7 6AA, UK}
\altaffiltext{9}{School of Physics and Astronomy, University of St Andrews, North Haugh, St Andrews KY16 9SS, UK}
\altaffiltext{10}{Center for Cosmology and Particle Physics, Department of Physics, New York University, 4 Washington Place, New York, NY 10003}
\altaffiltext{11}{Apache Point Observatory, P.O. Box 59, Sunspot, NM 88349, USA}
\altaffiltext{12}{Smithsonian Astrophysical Observatory, 60 Garden Street, Cambridge, MA 02138, USA}
\altaffiltext{13}{Instituto de Astronomia, Universidad Nacional Autonoma de Mexico, A.P. 70-264, 04510 Mexico D.F., Mexico}

\begin{abstract}

MaNGA (Mapping Nearby Galaxies at Apache Point Observatory) is an integral-field spectroscopic survey that is one of three core programs in the fourth-generation Sloan Digital Sky Survey (SDSS-IV).  MaNGA's 17 pluggable optical fiber-bundle integral field units (IFUs) will observe a sample of 10,000 nearby galaxies distributed throughout the SDSS imaging footprint (focusing particularly on the North Galactic Cap).  In each pointing these IFUs are deployed across a 3$^{\circ}$ field; they yield spectral coverage 3600-10,300 \AA\ at a typical resolution $R \sim 2000$, and sample the sky with 2'' diameter fiber apertures with a total bundle fill factor of 56\%. Observing over such a large field and range of wavelengths is particularly challenging for obtaining uniform and integral spatial coverage and resolution at all wavelengths and across each entire fiber array.  Data quality is affected by the IFU construction technique, chromatic and field differential refraction, the adopted dithering strategy, and many other effects. We use numerical simulations to constrain the hardware design and observing strategy for the survey with the aim of ensuring consistent data quality that meets the survey science requirements while permitting maximum observational flexibility.  We find that MaNGA science goals are best achieved with IFUs composed of a regular hexagonal grid of optical fibers with rms displacement of 5 $\micron$ or less from their nominal packing position; this goal is met by the MaNGA hardware, which achieves 3 $\micron$ rms fiber placement.  We further show that  MaNGA observations are best obtained in sets of three 15-minute exposures dithered along the vertices of a $1.44$ arcsec equilateral triangle; these sets form the minimum observational unit, and are repeated as needed to achieve a combined signal-to-noise ratio of 5 \AA$^{-1}$ per fiber in the $r$-band continuum at a surface brightness of 23 AB arcsec$^{-2}$.  In order to ensure uniform coverage and delivered image quality, we require that  the exposures in a given set be obtained within a 60 minute interval  of each other in hour angle, and that all exposures be obtained at airmass $\lesssim 1.2$ (i.e., within 1-3 hours of transit depending on the declination of a given field).

\end{abstract}



\section{INTRODUCTION}

Integral field spectroscopy (IFS) at optical and infrared wavelengths
is among the most significant developments in modern observations of
galaxies at all redshifts because it combines the benefits of
two-dimensional photometric analysis with physical diagnostics of
baryon composition and kinematics \citep[e.g.,][]{emsellem04, law09,
  bershady10, sanchez12, weijmans14, fabricius14}.  Recent advances
now enable multi-object IFS with instruments such as SAMI
\citep{croom12}, KMOS \citep{sharples13}, and MaNGA \citep{drory15}.
As a part of the 4th generation of the Sloan Digital Sky Survey
(SDSS-IV), the MaNGA (Mapping Nearby Galaxies at APO) project
\citep{bundy15} bundles fibers from the BOSS (Baryon Oscillation Spectroscopic Survey)
spectrograph \citep{smee13} into integral-field units to obtain
spatially resolved optical spectroscopy of 10,000 nearby galaxies over
a 6 year survey. Early results obtained with prototype MaNGA hardware
\citep{belfiore15, li15, wilkinson15} demonstrate the richness of the
data for exploring the stellar and gas composition.

Because current large-format detectors lack energy resolution
throughout most of the electromagnetic spectrum, IFS has adopted a
range of technical approaches to down-selecting and formatting a subset
of the three-dimensional data cube of wavelength and spatial position
onto a two-dimensional detector array. These approaches yield
different, science-driven trades in the data-cube
sampling. Simultaneous and integral coverage of the spatial field is
desirable and achieved by a number of instruments using lenslets
\citep[e.g., SAURON, OSIRIS;][]{bacon01,larkin03} or image slicers
\citep[e.g., SINFONI, MUSE;][]{eisenhauer03,bacon10}. However, the two
current wide-field, multi-object, IFS instruments--SAMI and MaNGA--use
bare-fiber arrays to minimize cost while maximizing
flexibility and patrol area, but at the penalty of not achieving truly integral
spatial coverage at any one time. This shortfall can be overcome by
careful attention to the interplay of the hardware design of the fiber
bundles and the observing strategy. 

The most immediate challenge is that the MaNGA fiber bundle, composed
of circular apertures with large interstitial gaps that significantly
undersample the PSF at the focal plane of the telescope, has a
non-uniform response across each IFU.  This means that (under
most techniques for the reconstruction of images from the data) the appearance
of objects that are small with respect to the fiber size (e.g., AGN or
\htwo\ regions) can vary across an IFU.  The reconstructed image of
such unresolved objects can either look small and circular (if the
object was centered on a single fiber), large and circular (if the
object was centered in the interstitial gap between three fibers),
highly elongated (if the object was centered midway between two
fibers), along with any range of shapes in between.

This is highly undesirable from a science standpoint, and therefore
typical fiber-bundle IFU surveys \citep[e.g.,][]{croom12,sanchez12}
dither their observations. Small dithers of a fraction of the fiber
spacing sample the missing points in the image plane and allow
reconstructed images based on multiple, dithered exposures to achieve
fairly uniform and integral spatial coverage.

This dithering is complicated by atmospheric refraction however,
especially given the
extremely wide spatial and spectral coverage of MaNGA.  Chromatic
differential refraction over the MaNGA wavelength range
($\lambda\lambda\ 3600-10300$ \AA) can be comparable to the diameter
of individual fibers, and field differential refraction (from
variation in the amount and direction of refraction over the
$3^{\circ}$ field of an SDSS plugplate) contributes similarly.  These
effects combine to degrade the effectiveness of a regular dithering
scheme in sampling the image plane.

This paper presents  simulations that explore the impact of
these effects on the expected MaNGA data quality, and thereby
constrain the hardware design and observing strategy for the
survey. In \S \ref{hardware.sec} we give an overview of the SDSS 2.5m
telescope and plugplate system, along with a brief description of the
MaNGA legacy hardware and IFU ferrule designs considered for the
survey.  We describe the basic design considerations for the survey in
\S \ref{basics.sec}.  Using the science requirements summarized in \S
\ref{reqperformance.sec}, typical integration times set by the read
noise characteristics of our detectors (\S \ref{itime.sec}), and
numerical simulations (\S \ref{simulations.sec}) we motivate the need
for dithered observations and regular hexagonal packing of the IFU
fiber bundles, culminating in a baseline hardware design and observing
strategy described in \S \ref{baseline-strat.sec}.  This baseline
observing strategy is significantly complicated by atmospheric
differential refraction, and we discuss the impact of chromatic and
field differential refraction on our data quality in \S
\ref{ar.chromatic.sec} and \ref{ar.field.sec} respectively, defining a
uniformity statistic $\Omega$ to describe the data quality in \S
\ref{omega.sec}.  Using the $\Omega$ statistic we formulate our final
observing strategy in terms of visibility windows in \S
\ref{strategy.sec}, noting a few additional practical considerations
(e.g., dithering accuracy and IFU bundle rotation) in \S
\ref{additional.sec}. We summarize our conclusions in \S
\ref{summary.sec}.



\section{Observatory and Hardware Overview}
\label{hardware.sec}

\subsection{Observatory and Legacy Hardware}

MaNGA operates on the SDSS 2.5-m telescope \citep{gunn06} located at Apache
Point Observatory (APO; latitude $\phi = +32^{\circ} \, 46' \, 49''$).
The telescope is a modified Ritchey-Chretien with alt-az mount that is designed with an interchangeable
cartridge system that can be installed at the Cassegrain focus.
The MaNGA hardware is described in greater detail by \citet{drory15}; here we briefly review
the major salient features of the system.

MaNGA has 6 cartridges, each of which contains
a plugplate with a field of view $\sim 3^{\circ}$ in diameter that has been pre-drilled with holes corresponding to the locations of target galaxies into which 
optical fibers and IFUs can be plugged each day in preparation for a night of observing.  
These plates are fixed at zero degrees position angle
(i.e., the on-sky orientation of the telescope focal plane coordinate reference frame is fixed).  

Each MaNGA cartridge has a total of 1423 fibers (709 on spectrograph 1, 714 on spectrograph 2),
corresponding to 17 science IFUs ranging in size from 19 to 127 fibers ($12.5 -32.5$ arcsec diameter; 1247 fibers total),
twelve 7-fiber mini-bundles used for spectrophotometic calibration (84 fibers total; see Yan et al. in prep), and
92 single fibers used for sky subtraction
that can be deployed within a 14' radius of their associated IFU harness.\footnote{The physical size of the hardware components also defines a
minimum-distance exclusion zone around each plugged object.  These exclusion distances are 116'' (7 mm), 89'' (5.35 mm), and 62'' (3.7 mm) for
IFU-IFU, IFU-sky, and sky-sky fiber placement respectively.}
Each IFU has its
rotation fixed using alignment pins in the ferrules that plug into corresponding alignment holes located a short
distance West of each target galaxy. 

These optical fibers feed the twin BOSS \citep[Baryonic Oscillation Spectroscopic Survey,]{dawson13}
spectrographs \citep{smee13}.
The collimated beams in each spectrograph are split with a dichroic and 
feed a blue ($\lambda\lambda 3600-6000$ \AA) and red camera ($\lambda\lambda 6000 - 10300$ \AA).
The blue cameras use blue-sensitive 4k $\times$ 4k e2V CCDs while the red cameras
use 4k $\times$ 4k fully-depleted LBNL CCDs; all cameras have 15$\micron$ pixels.
Spectral resolution varies with wavelength from $R = \lambda/\delta\lambda \sim 1400$ at 3600 \AA\ to $R \sim 2000$ at 6000 \AA\ (blue channel),
and $R \sim 1800$ at 6000 \AA\ to $R \sim 2200$ at 10300 \AA\ \citep[red channel; see Fig. 36 of][]{smee13}.
Spectra from each of these four cameras are extracted and processed through sky subtraction,
spectrophotometric calibration, astrometric registration, and reconstructed into three-dimensional
data cubes using a software pipeline (Law et al. in prep)
descended from that previously used for BOSS
\citep[idlspec2d; see][Schlegel et al. in prep]{bolton12}

The telescope guider system is optimized for a wavelength of $\sim$ 5500 \AA\ and uses endoscopic fibers inserted into 16 holes in each plugplate
corresponding to the locations of bright guide stars.  These endoscopic fibers produce images of the guide stars
on a guider camera, and the guider actively adjusts the focus, scale, rotation,
and offset of the telescope focal plane to track these stars through varying weather conditions and observing angles.


\subsection{IFU Ferrule Design}
\label{ferrules.sec}

The ability of an IFU fiber bundle to deliver good, repeatable, and uniform image quality depends
most fundamentally on the arrangement of fibers within the bundle; while dithering (\S \ref{dithering.sec}), differential refraction (\S \ref{ar.sec}),
and other considerations are important, the fiber placement sets the basis
for the sampling regularity of the entire survey.

\begin{figure}
\epsscale{1.2}
\plotone{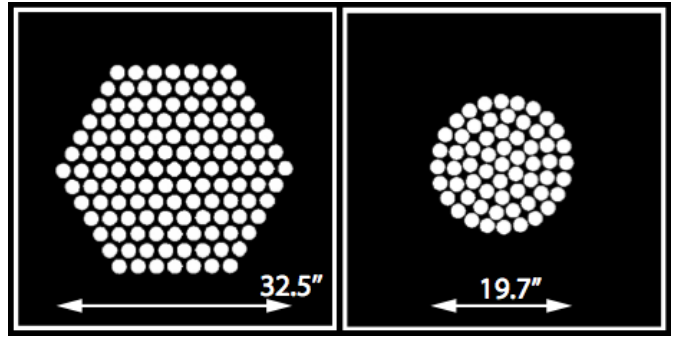}
\caption{Fiber bundle designs considered for MaNGA (white regions represent live fiber cores).  The 
left-hand panel shows a 127-fiber bundle for which the fibers are arranged in a regular hexagonal array (i.e., the final MaNGA IFU design; shown here is
as-built harness ma024); the right-hand panel shows an example bundle of 61 fibers in a circular packing arrangement 
based on that adopted by the SAMI team for use at the Australian Astronomical Observatory \citep[][compare their Fig. 3]{croom12}.
Although the hexagonal arrangement of fibers has greater regularity, the circular arrangement has greater effective filling factor since the protective
buffers are stripped.}
\label{bundlesims0.fig}
\end{figure}

As described by \citet{drory15}, the MaNGA fibers have an inner light-sensitive core diameter (ID) of 120 $\micron$
(corresponding to 2.0 arcsec in the telescope focal plane)
and an outer diameter (OD) of $151.0 \pm 0.5$ $\micron$ with their protective buffers and cladding.
We originally considered two kinds of fiber bundles for MaNGA, as illustrated in Figure \ref{bundlesims0.fig}.  The first was a circular bundle of fibers 
that maximizes the filling factor
of light-sensitive fiber cores relative to the total IFU footprint by chemically stripping the protective buffers from the ends of each fiber.
As developed for the SAMI survey by \citet{bland-hawthorn10},
these `Sydney-style' bundles maximize
the effective filling factor at the cost of decreased fiber throughput due to focal ratio degradation (FRD), greater fragility of the glass cores, and irregular fiber packing due to the circular ferrule geometry.
Based on the numerical performance simulations described in \S \ref{regularity.sec}, we  prototyped
(and ultimately chose to adopt)
a second style of fiber bundle composed of a regular arrangement of buffered fibers within a tapered hexagonal ferrule
for which we pioneered a novel construction technique  \citep[see details in][]{drory15}.  While reaching lower effective filling factor, this technique
improves fiber throughput,\footnote{A conservative estimate can be made by comparing Figure 4 of \citet{croom12} to Figure 11 of \citet{drory15}: MaNGA achieves 
95 $\pm1$ \% throughput with an exit f-ratio of f/4 for fibers fed at f/5. In contrast, the original SAMI bundles achieved 50-75\% throughput with an exit f-ratio of f/3.15 fed at f/3.4.  
We note that the FRD of even the second-generation SAMI bundles \citep[Fig. 5 of][]{bryant14} is sufficiently large that it would require our optics to be 40\% larger in area to collect
the same ensquared energy given the Sloan telescope feed.}
decreases breakage,\footnote{After $\sim$ 6 months of operation, 7 individual fibers within IFUs have broken (1 in manufacturing, 1 in assembly, 5 in operation), representing $< 0.1$\% of the total.
Detailed statistics on the breakage frequency of stripped, fused fiber bundles are unknown but would have represented a significant cost increase in manufacturing.} 
and (by virtue of its hexagonal geometry) 
permits extremely regular fiber placement within each IFU.

The theoretical effective fiber packing density of the hexagonal IFUs can be defined as the ratio of the total fiber
core area ($A_{\rm core}$) to the area of the hexagon circumscribing the fiber bundle ($A_{\rm hex}$), where:

\begin{equation}
A_{\rm hex} = \frac{\sqrt{3}}{2} d^2 (\sqrt{3} N_{\rm R} +1)^2
\end{equation}

\begin{equation}
A_{\rm core} = \pi \left( \frac{d-2 t}{2}\right)^2 (1 + 3 N_{\rm R}(N_{\rm R}+1))
\end{equation}

\begin{figure}
\plotone{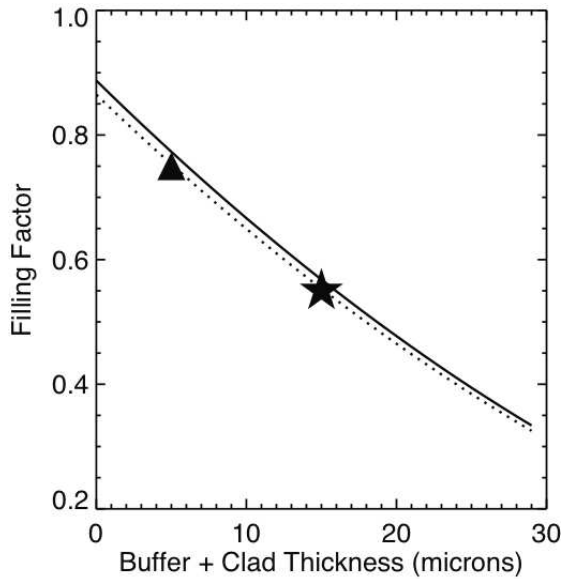}
\caption{Effective IFU filling factor (live fiber core area divided by total IFU footprint)
as a function of buffer thickness for an ideal 127-fiber (solid line) and a 19-fiber (dotted line) hexagonal IFU.  The small difference between the solid and dotted lines
represents the diminishing importance of edge effects in the hexagonal footprint as the IFU area increases.  The filled star represents the measured 56\%
filling factor of the as-built 127-fiber MaNGA IFUs   \citep{drory15}, which is consistent with theoretical expectations.  The filled triangle shows the 75\% filling factor
of the SAMI survey bundles (5 $\micron$ cladding) for comparison.}
\label{fillfac.fig}
\end{figure}

Here $d = 151$ \micron\ is the outer diameter of an individual fiber, $t = 15.5$ \micron\ is the thickness of the fiber buffer and cladding, and $N_{\rm R}$ is the 
number of `rings' in the bundle ($N_{\rm R} = 2$ for a 19-fiber IFU, and $N_{\rm R} = 6$ for a 127-fiber IFU).
In Figure \ref{fillfac.fig} we plot the effective filling factor $f = A_{\rm core}/A_{\rm hex}$ as a function of the buffer thickness $t$.  In accord with these predictions, 
the prototype circular Sydney-style bundles (whose fibers are chemically etched to an outer diameter of $\sim 132$ \micron) achieve a filling factor of $\sim 70$\%, while the 
as-built hexagonal bundles with fully-buffered fibers achieve a filling factor of 56\%.

\section{Basic Considerations}
\label{basics.sec}

\subsection{Required Performance}
\label{reqperformance.sec}

Since the fiber bundles consist of  2'' diameter circular apertures separated by large interstitial gaps, each exposure will 
significantly undersample the point spread function (PSF) at the focal plane of the telescope
(typically $\sim 1.5$'') and produce a non-uniform
response function across the face of each IFU. 
We require that the MaNGA IFUs deliver sufficiently uniform performance that physical structures do not vary in shape as
a function of where they happen to fall within the IFU (i.e., a circular star forming region within a galaxy should appear circular in the final MaNGA data cube
regardless of whether it is in the center or the outskirts of the galaxy).

A convenient way to place a limit on the level of uniformity required is to ensure that variations in the 2d PSF of the reconstructed MaNGA data cubes do not
significantly impact measurements of the Balmer decrement or BPT-style \citep[e.g.,][]{baldwin81} line ratio diagrams.
Since atmospheric differential refraction shifts the effective position of each fiber as a function of wavelength (\S \ref{ar.sec}), \otwo\ and \Ha\ observations of
a given \htwo\ region for instance will be obtained with a slightly {\it different} configuration of fibers -- 
while \Ha\ emission may be centered in a given fiber, spatially coincident
\otwo\ emission may be centered in the interstitial region between fibers.

As outlined in the MaNGA Science Requirements Document (SRD; see Yan et al. in prep), relative spectophotometry 
between \otwo\ ($\lambda = 3727$ \AA) and H$\alpha$ ($\lambda = 6564$ \AA) must be accurate to 7\% or better in order to obtain
the desired constraints on the star formation rate (SFR) and nebular metallicity
within galaxies.  We therefore explore how this required spectrophotometric accuracy translates to limits on the spatial variability of the MaNGA
PSF.

\begin{figure}
\plotone{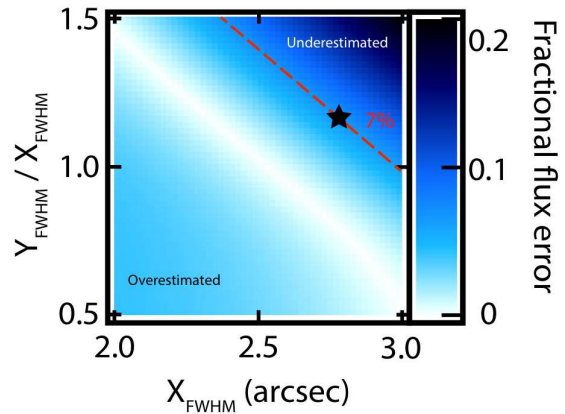}
\caption{Fractional error in the recovered flux from a point source if the assumed FWHM and axis ratio were incorrect.  The reference source
is taken to have a circular gaussian PSF with FWHM 2.5 arcsec; integrating the flux within a 2.5 RMS width aperture 
nominally encloses 95.7\% of the total
flux.  If the actual FWHM is smaller (larger) than the model along any dimension the total flux enclosed by the aperture increases (decreases), 
resulting in an
overestimate (underestimate) of the total flux.  The dashed red line indicates the 7\% error threshhold set by the MaNGA SRD; the solid black
star indicates our adopted limits on the allowable variability of the delivered MaNGA PSF (15\% in axis ratio, and 15\% in circularly averaged
PSF FWHM). }
\label{fluxfig.fig}
\end{figure}

We begin by assuming that the PSF in a typical MaNGA reconstructed data cube
can be characterized by a circular gaussian with a FWHM of 2.5 arcsec
(as we discuss at greater length in Law et al., in prep, this model is a good approximation to the MaNGA commissioning data).
Using typical aperture photometry techniques, a circular aperture of radius 
$2.66$ arcsec (i.e., 2.5 times the radial scalelength of the PSF) would nominally
enclose 96\% of the total flux.\footnote{If we were to adopt a PSF model with more power in the wings, or shrink the size of the
circular aperture the variability between different PSF shapes would increase and lead to more stringent constraints on the 
allowable variability in the delivery MaNGA PSF.}
In Figure \ref{fluxfig.fig} we illustrate how deviations from the nominal PSF model
would affect this total; as the PSF becomes broader or more elongated the flux contained within the fixed aperture decreases,
meaning that the derived aperture-corrected total fluxes would be in error.\footnote{If the goal were to measure the 
flux from a single bright source whose structure is known a-priori to be effectively a point source then
the actual light profile could be measured at each wavelength and the aperture adjusted accordingly.  However, such
a-priori knowledge of the intrinsic source structure cannot generally be assumed.  Similarly, we assume that wavelength-dependant variations from the $\lambda^{-1/5}$ Kolmogorov atmospheric turbulence profile are taken into account in determining the appropriate aperture.}
In particular, we find that an error of 20\% in the profile FWHM and 15\% in the profile minor/major axis ratio is sufficient to bias
the resulting flux measurements at the 7\% level (filled star in Figure \ref{fluxfig.fig}).

Similarly, in order to ensure that our limiting fluxes for {\it undetected} nebular transition features are accurate at the 7\% level we also require
that the signal-to-noise ratio of our data is constant at the 7\% level across each IFU.  Since the limiting flux is proportional to the 
square root of the exposure time, this translates
to a requirement that the exposure time is effectively constant across each IFU at the 15\% level.

These three metrics (circularity, FWHM, and signal-to-noise ratio) therefore set our 
{\it requirements} on the uniformity of the reconstructed image profile such that the calibrated fluxes derived from MaNGA data
cubes are accurate at the 7\% level.  Ideally, however, we would prefer that spatial sampling issues not dominate the
flux calibration accuracy budget for the MaNGA data cubes, and we therefore set a {\it goal} of achieving photometric 
performance at the 3.5\% level where possible.  The MaNGA hardware construction, dithering pattern, and observing strategy is 
therefore set by the following three high-level considerations:

\begin{enumerate}
\item The reconstructed FWHM of all angular resolution elements in a bundle should vary by $<$ 10\% (goal) or 20\% (requirement) across each IFU.
\item The reconstructed minor-to-major axis ratio of all resolution elements in a bundle should be $b/a \geq 0.93$ (goal) or 0.85 (requirement) across each IFU.
\item The effective integration time of all resolution elements in a bundle should vary by $<$ 7\% (goal) or 15\% (requirement) across each IFU.
\end{enumerate}


\subsection{Integration Time}
\label{itime.sec}

The total integration time is set by our requirement that MaNGA reach a 
signal-to-noise ratio of 5 \AA$^{-1}$ fiber$^{-1}$ in the $r$-band continuum  at a surface brightness of 23 AB arcsec$^{-2}$.
As described by Wake et al. (in prep.) and Yan et al. (in prep.) the typical integration time per plate to reach this target
is anticipated
to be about 3 hours in median conditions.  In good conditions however the required time could be as low
as 1.5-2 hours, and for particularly low-latitude fields the required time could be as much as 4-5 hours.
This substantial variation in total exposure time requires an observing strategy flexible enough to
accommodate it.

The optimal integration time for individual exposures is constrained by the MaNGA hardware
and typical background sky spectrum at APO.  One of the strengths of MaNGA is the high throughput of the
BOSS spectrographs shortward of 4000 \AA, and we therefore integrate each exposure
for long enough that the shot noise from
the background sky spectrum and detector dark current exceeds the read noise.  The total noise $N$ 
as a function of wavelength is given by
\begin{equation}
N (\lambda) = \sqrt{(f_{\rm s}(\lambda) +f_{\rm d} \, n_1) \, t + n_1 \, N_{\rm r}^2 }
\label{noise.eqn}
\end{equation}
where $f_{\rm s}(\lambda)$ is the background sky spectrum in units of e$^-$ minute$^{-1}$ per spectral pixel, $f_{\rm d}$ is the dark current in e$^-$ pixel$^{-1}$ minute$^{-1}$, $N_{\rm r}$ is the read noise in e$^-$ pixel$^{-1}$, $n_1 = 3$ pixels is the spatial width of a spectrum on the detector (see discussion by Law et al. in prep), 
and $t$ is the integration time of an exposure in minutes.
Rearranging Eqn. \ref{noise.eqn} we find the time $t_{\rm min}$ required for the combined sky background and dark current  to equal the read noise:

\begin{equation}
t_{\rm min} (\lambda) = \frac{ n_1 \, N_{\rm r}^2}{f_{\rm s}(\lambda) + f_{\rm d} n_1}
\end{equation}

We estimate $f_{\rm s}(\lambda)$ for a typical MaNGA dark-time 
observation using commissioning data from all-sky plate 7341
(i.e., a calibration plate for which all IFUs target regions of blank sky)
observed on MJD 56693.\footnote{MJD (Modified Julian Date) 56693 corresponds to
February 5, 2014.}
Following the data model outlined by Law et al. (in prep), we take
the FLUX array of the reduced mgFrame file (in units of flatfielded e$^{-}$ per spectral pixel), multiply
by the SUPERFLAT array to obtain spectra in raw $e^{-}$ per spectral pixel, and combine $\sim 600$  
individual fiber spectra to construct an extremely high-precision model of the background sky.
We take the detector read noise to be $R_{\rm n} = 2.0$ (2.8) e$^-$ pixel$^{-1}$, and the dark current
to be 0.033 (0.066) 
e$^-$ pixel$^{-1}$ minute$^{-1}$
for the blue (red) camera  \citep[see Table 4 of][]{smee13}.\footnote{The dark current is typically $\lesssim 2$\% of the dark-time
sky background signal.}

\begin{figure*}
\plotone{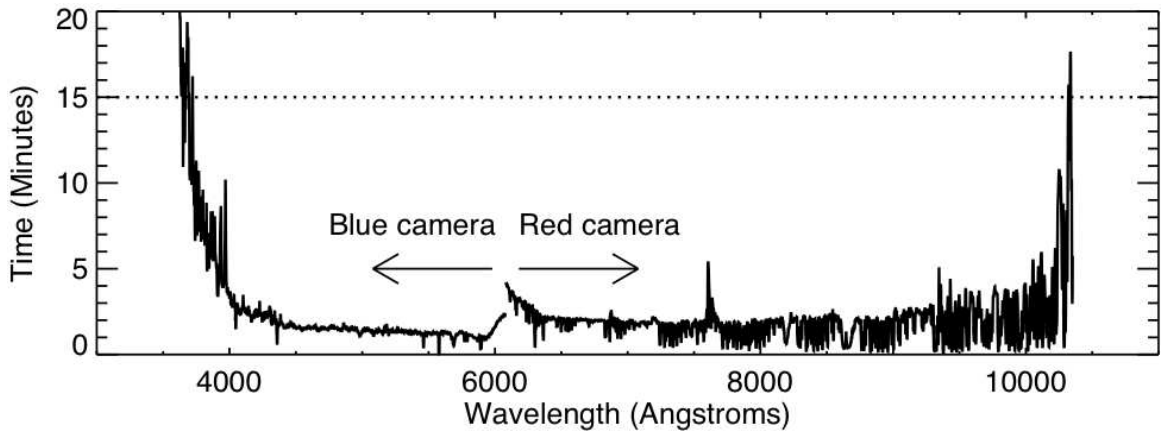}
\caption{Exposure time $t_{\rm min}$ required for a typical MaNGA dark-time sky spectrum to be dominated by Poisson noise from the background sky plus detector dark current.  
The break around 6000 \AA\ represents the dichroic
break between red and blue channels; in reality there is a $\sim 300$ \AA\ overlap between these channels.  Strong features longwards of $\sim 8000$ \AA\
are due to bright OH sky lines.
The dashed line indicates the adopted 15 minute exposure time.}
\label{inttime.fig}
\end{figure*}

We plot $t_{\rm min}$ as a function of wavelength in Figure \ref{inttime.fig}, and note that the sky background rapidly
dominates over read noise at almost all wavelengths, especially in the vicinity of strong OH atmospheric emission
lines in the near-IR.  The upturn in $t_{\rm min}$ shortward of 4000 \AA\ represents the falloff in blue sensitivity of
the detectors, but an integration time of 15 minutes per exposure ensures that observations are shot-noise dominated
for all $\lambda > 3700$ \AA.

Although an integration time of longer than 15 minutes would further decrease the contribution of read noise
to the total error budget, such longer integrations are undesirable because of the cosmic ray event rate recorded by the red-channel detectors.
In practice, the maximum integration time of each exposure is also limited by differential atmospheric refraction considerations (see \S \ref{omegaexp.sec}),
and we therefore adopt a nominal time of 15 minutes per exposure.  Each completed plate will therefore consist of $\sim 6 - 20$ exposures
in order to reach the target depth.


\subsection{Numerical Simulations}
\label{simulations.sec}

\subsubsection{Simulation Method}
\label{simmethod.sec}

In order to assess the relative performance of different IFU bundles and observing techniques we
perform a series of numerical simulations 
designed to test the uniformity of their response to unresolved point sources (for which spatial structure is most pronounced).
Adopting a working box size of $\sim 45 \times 45$ arcsec with simulated pixels spaced every $0.1$ arcsec we first compute the footprint of a given IFU; 
this defines a mask image for which each fiber in the IFU
is associated with a given set of pixels
in the telescope focal plane that its light-sensitive core subtends.  We then create an input `image' to be observed
by the simulated MaNGA IFUs by convolving a delta function by a
model of the PSF at the focal plane of the SDSS 2.5-m telescope.
This focal-plane PSF is taken to be the sum of two Gaussian profiles with 
FWHM $\theta$ and $2\theta$ respectively (where $\theta=1.4$ arcsec is 
the FWHM of the median atmospheric seeing
profile divided by 1.05)
and peak amplitude ratio of 9/1.\footnote{Mathematically, this is equivalent to the linear sum
of 9/13 times the input image convolved with a Gaussian of FWHM $\theta$ plus 4/13 times the input image
convolved with a Gaussian of FWHM $2 \theta$.  This profile provides a reasonable approximation
of the on-axis SDSS focal plane PSF, matching the inner parts of the profile well and accounting for most
of the flux in the outer wings (J.E. Gunn, priv. comm.).}
This input image is convolved with the top-hat fiber mask 
to determine the total amount of light received by each fiber; although the present simulation considers only a single input image the technique
is immediately generalizable to multi-wavelength input image slices.

We reconstruct a two-dimensional image from the individual fiber fluxes using a flux-conserving variant
of Shepards method (inverse-distance weighting) similar to that used by the CALIFA survey \citep{sanchez12}.  As part of MaNGA design simulations we explored alternative
methods of image reconstruction such as drizzling \citep[e.g., as adopted by SAMI, see][]{sharp15}, 
thin-plate-spline fits, minimum curvature surface fits, and kriging.  As discussed by Law et al. (in prep) the modified Shepard's method yielded the best results,
and here we adopt the same parameters (e.g., final spaxel scale of 0.5 arcsec)
as used by the MaNGA Data Reduction Pipeline (DRP) for genuine survey data.
The reconstructed image is fit with a 2d Gaussian model to determine its FWHM
and axial ratio; major axis rotation is left as a free parameter.

\begin{figure*}
\plotone{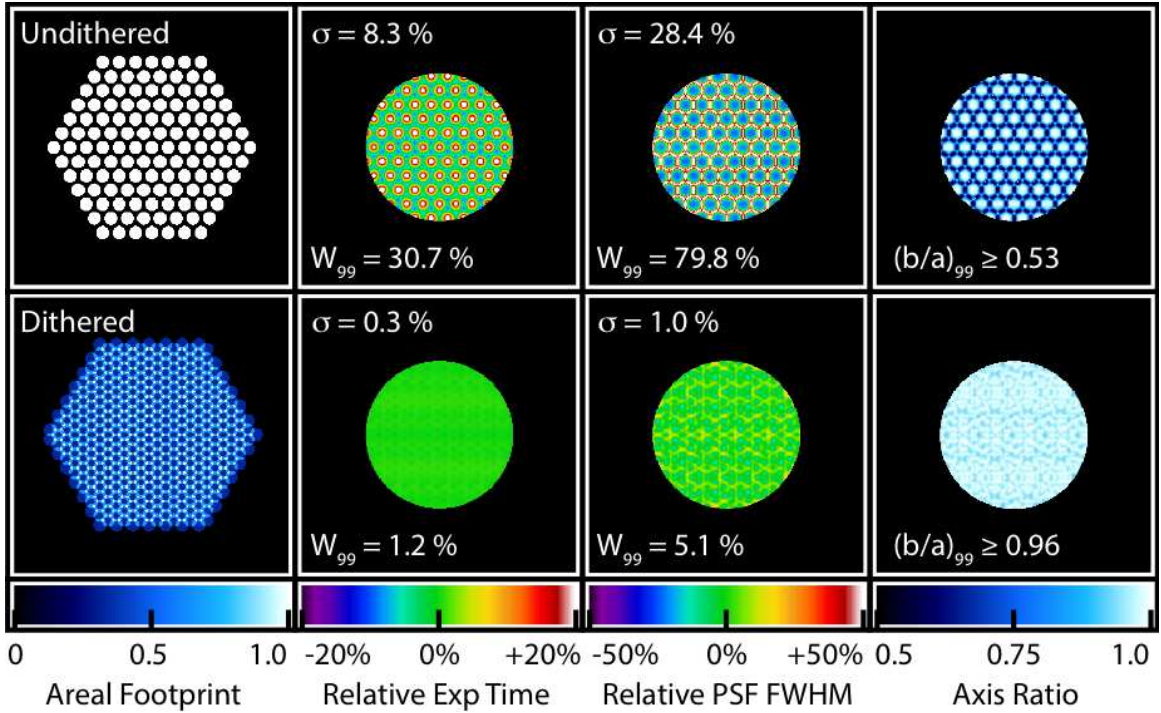}
\caption{Simulations of point-source response as a function of location within an IFU for a single exposure (top row) and a dithered set of exposures (bottom row) 
using a theoretically perfect hexagonal fiber bundle.  The left-most panels show the footprint of the IFU fibers on the sky,
the second column of panels show the percentage variations about the median exposure time as a function of position within the bundle.  The third column
of panels shows the deviation from the 
median delivered PSF, and the right-hand column of panels shows 
the recovered minor/major axis ratio.  For undithered observations the greatest effective depth is obtained for sources located in the middle of a fiber (as is the
smallest and most circular reconstructed image of a point source), while point sources falling in interstitial regions between fibers have minor/major axis ratios as low as $\sim 0.5$ and
FWHM nearly double that of sources centered within a fiber.  Numbers in panels 2-4 indicate the RMS deviation between values ($\sigma$), the $3\sigma$ width encompassing 99\% of all values
($W_{99}$), and the minor/major axis ratio above which 99\% of point lie ($(b/a)_{99}$).}
\label{bundlesims1.fig}
\end{figure*}

This exercise is repeated for delta functions located in each of the 0.1 arcsec grid squares that lie within the central 75\% of the IFU fiber bundle footprint
(i.e., ignoring edge effects from point sources located on the outer ring of an IFU), resulting
in $\sim 40,000$ simulated points across a 127-fiber IFU bundle.
In Figure \ref{bundlesims1.fig} (top row) we plot the on-sky footprint of a hexagonal fiber array, along with the variations in effective exposure time
(exposure time multiplied by the fraction of the total light that is collected by fibers rather than being lost to interstitial regions), 
FWHM, and minor-to-major axis ratio of the reconstructed point spread function (PSF) as a function of the location of the point source within the fiber bundle.  
As anticipated, we note that all three quantities vary substantially across a given IFU in a single exposure.

We quantify these results by calculating the RMS of the distributions in effective exposure time and reconstructed PSF FWHM (relative to the median
values as $[X - X_{\rm median}]/X_{\rm median}$),
the $3\sigma$ width $W_{99}$ encompassing 99\% of these values, and the 99\% lower bound for the minor-to-major axis ratio.
For a single exposure, the effective integration time varies by $W_{99} = 30.7$\% around the 
median\footnote{The median effective exposure time is just the filling factor (0.56) times the actual exposure time.} value;
unsurprisingly, the greatest fraction of the total light is recorded for objects that are centered in a fiber, while the least amount is recorded
for objects in interstitial regions.  
Similarly, the reconstructed PSF FWHM varies by almost 80\% (from $\sim$ 2 to 4 arcsec) depending on where a source falls with respect to the fiber grid,
and the minor-to-major axis ratio $b/a$ of the reconstructed image varies from $\sim 0.5 - 1.0$ (99\% of values $b/a \geq 0.53$).

In practical terms, this means that an unresolved \htwo\ region observed with such an IFU for just a single exposure may appear to be compact and circularly symmetric if it lands
directly in the middle of a fiber, elongated and skinny if it falls directly between two fibers, or large and triangular if it falls midway between three
adjacent fibers.\footnote{Strictly, a single exposure simply does not have the spatial sampling in these cases to discriminate (for instance) between an unresolved point
source and an elongated source.}
Allowing for the effects of chromatic differential refraction (\S \ref{ar.chromatic.sec}), this means that a 
single such \htwo\ region may simultaneously be sampled by {\it all three} different such configurations at different wavelengths.


\subsubsection{Dithering}
\label{dithering.sec}

\begin{figure*}
\plotone{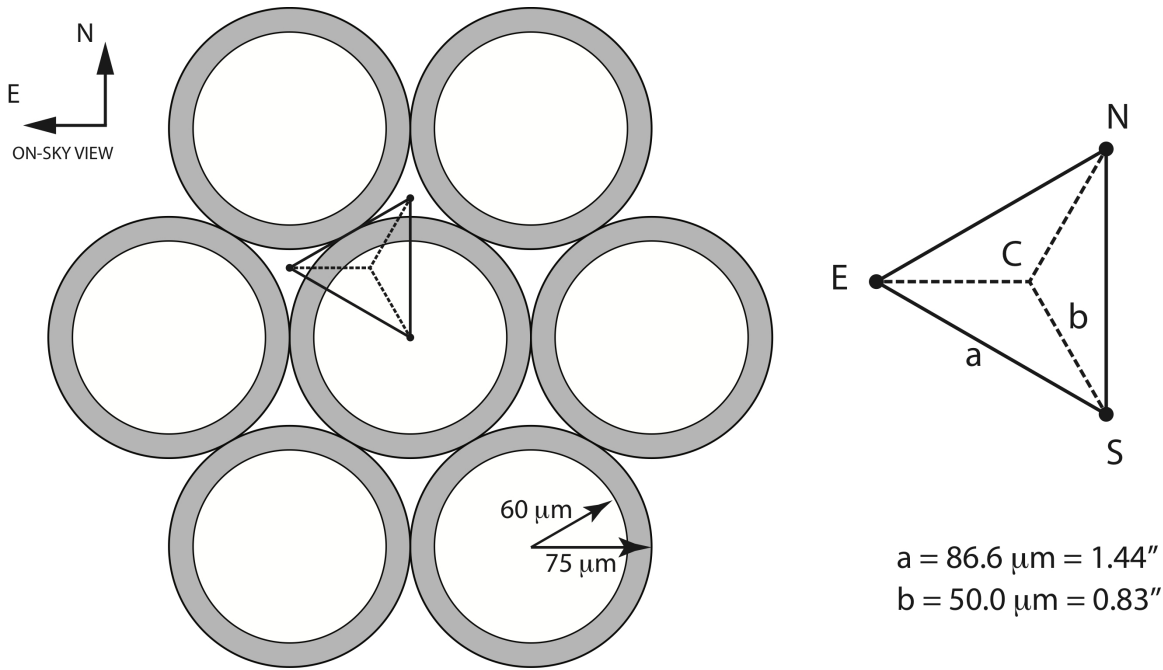}
\caption{Schematic diagram of the 7 central fibers within a hexagonally-packed MaNGA IFU, showing the 120 micron diameter fiber core
and surrounding cladding plus buffer.  The triangular figure shows the relative positions of the three dither positions; the fiber
bundle is located at position `S'.
The central (C) `home' position is labeled, along with the north (N), south (S), and east (E) dither positions.  The nominal plate scale of the SDSS telescope
is 217.7358 mm/degree, or 60.48 microns/arcsec.}
\label{dither.fig}
\end{figure*}

The sampling irregularities from  fiber-bundle IFUs with substantial
interstitial light losses are well known from previous IFU surveys
 \citep[e.g.,][]{sanchez12,sharp15}, and can be largely overcome by obtaining dithered observations.
The geometry of the hexagonal
fiber arrangement readily lends itself to a fixed triangular three-point dithering scheme that effectively fills the interstitial regions
as illustrated in Figure \ref{dither.fig}.
Repeating the simulations performed in \S \ref{simmethod.sec} with such dithered observations, we find that the
combined data from just three exposures is able to achieve remarkably uniform image quality at all locations within a single IFU
(Figure \ref{bundlesims1.fig}, bottom row).  In contrast to the undithered case, 3-point dithering delivers effective exposure time constant to within 0.3\% RMS,
FWHM of $2.69 \pm 0.01$ arcsec, and ellipticity $\leq 0.04$.  This uniformity easily meets the MaNGA science requirements described in \S \ref{reqperformance.sec}.

Logically, the 3-point dithering pattern could be expanded to a regular 9-point pattern that also provides uniform coverage of the interstitial gaps, but with a finer
sampling of the image plane.  Although simulations suggest that this could provide $\sim 10$\% improvement in the delivered PSF FWHM, such gains
were not realized on-sky in tests with the MaNGA prototype hardware.
This lack of improvement with respect to theoretical calculations is likely due to a confluence of numerous complicating factors, including
degradation of the nominal dithering pattern by atmospheric refraction (see \S \ref{ar.sec}), variations in fiber-to-fiber sensitivity,
and changes in seeing and  transparency conditions between exposures (see \S \ref{weather.sec}).


\subsubsection{Fiber Packing Regularity}
\label{regularity.sec}

The gains achievable with such dithering depend fundamentally on the uniformity of each IFU fiber bundle so that a single telescope offset can simultaneously
dither each of our 29 IFUs
(17 science and 12 calibration bundles)
across the $3^{\circ}$ field such that their fibers align with the interstitial gaps from the previous exposure.
If fibers are not located at regular positions within every IFU, the dithering will not be able to uniformly sample the image plane.
We explore the effect of fiber packing irregularity by
repeating our earlier simulations with the introduction of a random
perturbation to the position of each fiber in the simulated IFU fiber bundle, such that each fiber is slightly offset from its nominal
position by some distance drawn randomly from a Gaussian distribution with a given rms.  Each simulated IFU bundle is observed with a nominal 3-pt dither pattern as defined by Figure \ref{dither.fig}.
Additionally, we simulate the effect of observing the circular Sydney-style fiber bundle with a 7-point dither pattern (based on that adopted by the SAMI survey) that 
compensates for the irregular fiber placement with greater filling factor and a larger number of dithered sampling points.

\begin{figure*}
\plotone{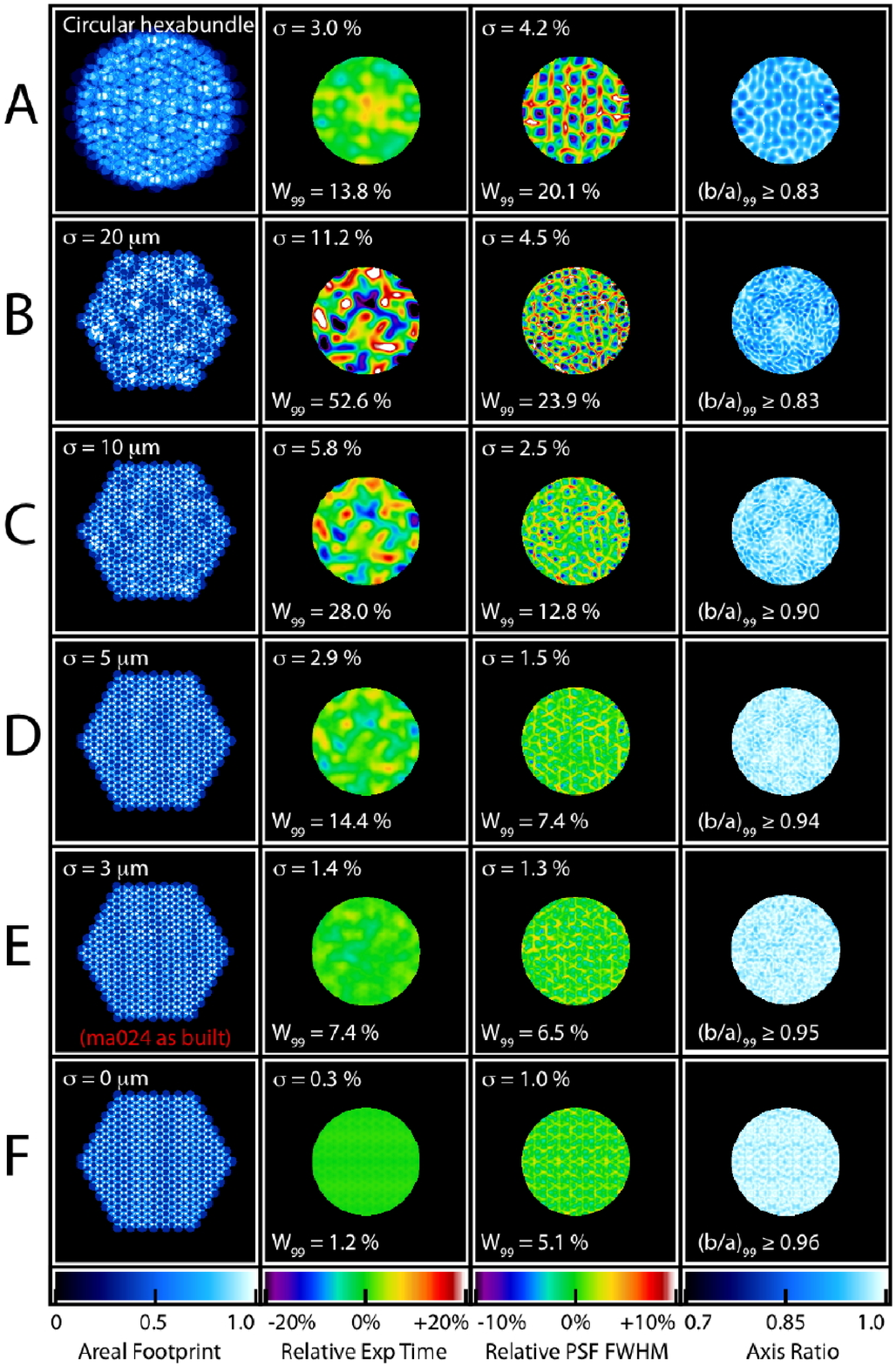}
\caption{As Figure \ref{bundlesims1.fig}, but showing simulated point-source response as a function of location in an IFU for dithered observations of fiber bundles built to a variety of specifications.
Row A simulates a Sydney-style 61-fiber bundle using a 7-point dither pattern.  Rows B-F simulate a 3-point dither pattern applied to a hexagonal arrangement of
127 fibers with varying RMS deviations of each fiber from the nominal position ($\sigma = 0 - 20$ \micron).  Note that for display purposes panel A is zoomed in slightly compared
to panels B-F.  In order to meet our uniformity criteria we require $\sigma < 5$ \micron, which our as-built IFUs achieve (row E).}
\label{bundlesims2.fig}
\end{figure*}

As indicated by  Figure \ref{bundlesims2.fig}, neither the dithered
Sydney-style circular fiber bundle nor the 20 $\mu$m tolerance hexagonal fiber bundles meet our target regularity goals, with
a recovered PSF FWHM\footnote{We quote the average of the minor- and major-axis FWHM values.} 
varying by $> 20$\% over the extent of an IFU (i.e., $2.66 \pm 0.12$ arcsec with 99\% values ranging from $\sim$ 2.3 - 2.9 arcsec), and minor/major axis ratios as low
as $b/a \sim 0.8$.  In contrast, using a hexagonal 
fiber array constructed to a tolerance of 5 \micron\ rms with a 3-point dither pattern we expect to achieve a PSF FWHM that varies by less than 10\% over
a given IFU.  

As detailed by  \citet{drory15}, the
as-built MaNGA fiber bundles meet and exceed our target threshold with a typical fiber placement accuracy of 3 \micron\ rms.
Using the as-measured fiber metrology\footnote{The final placement of individual fibers within an IFU can be measured to an accuracy of better than 1 \micron\  \citep{drory15}.}
for 127-fiber MaNGA bundle ma024, we simulate the anticipated performance using this fiber bundle in row E of Figure \ref{bundlesims2.fig}.  With a nominal dither pattern
we expect to achieve a PSF FWHM which varies by less 
than 7\% over an IFU (i.e., $2.66 \pm 0.01$ arcsec with 99\% values ranging from $\sim$ 2.6 to 2.7 arcsec), and has a nearly circular profile everywhere
with $b/a > 0.95$.


\subsection{Baseline Observing Strategy}
\label{baseline-strat.sec}

The dithered observing simulations presented in \S \ref{dithering.sec} and exposure time requirements described in \S \ref{itime.sec} motivate
a nominal observing scheme in which
targets are observed in  {\bf sets} of 3 dithered exposures (N-S-E) of 15 minutes each.
Given the regularity of fiber placement with each IFU
and the locator pins that constrain each IFU to have the same position angle, correctly-dithered exposures can be simultaneously obtained for
all IFUs on a given plate by simply offsetting the telescope pointing with respect to the guide stars.
Since the coverage and image quality of a single set of three dithered exposures is known to be acceptably uniform, the total summed coverage of $N$ such sets 
will also be uniform and have a depth of $0.75 N$ hours, allowing us to simply observe additional sets of 3 exposures
until the combined data reaches our target
signal-to-noise ratio of 5 \AA$^{-1}$ fiber$^{-1}$ in the $r$-band continuum  at a surface brightness of 23 AB arcsec$^{-2}$.

Such a scheme provides us with considerable flexibility to adjust our total exposure time in 45-minute increments without adversely impacting the delivered data quality
whether there are 6 or 20 total exposures for a given galaxy.
It is this flexibility as much as the dithered performance simulations themselves that drives us to adopt the regular hexagonal fiber arrays for MaNGA rather than the SAMI-style
circular fiber bundles, which rely upon a large number of exposures at many different dither positions to statistically fill in the interstitial gaps.\footnote{Additionally, the hexagonal 
tapered ferrule construction
technique can be scaled up to bundles with large numbers of hexagonal `rings' without significantly degrading the packing regularity.}
However, since this technique relies upon tightly controlling the fiber locations to provide uniform coverage we must properly mitigate a variety of effects that will act to degrade this
uniformity, and this goal in turn drives many aspects of the survey operation.


\section{Atmospheric Refraction}
\label{ar.sec}

As a photon passes through the Earth's atmosphere it is refracted by variations in the density of the air.
Under the usual assumption of a plane-parallel atmosphere with a vertical density gradient this bends the light from an astronomical target
along the parallactic angle (the great circle connecting the target and the observers local zenith), causing astronomical objects to appear slightly higher in
the sky than they truly are.
Atmospheric refraction introduces significant optical distortions that adversely affect our ability to 
dither our IFU observations to the  desired accuracy.  
Loosely speaking, the effects can be split into {\it chromatic differential refraction} and {\it field differential refraction} which we detail below.


\subsection{Chromatic Differential Refraction}
\label{ar.chromatic.sec}

Atmospheric refraction is a function of atmospheric conditions (temperature, pressure, and relative humidity), zenith distance (i.e., the amount of atmosphere that an incoming photon must traverse), and wavelength.
The impact of such refraction on astronomical observations has been studied at some length in the literature
\citep[e.g.,][and references therein]{filippenko82,cuby98}; we adopt estimates of the magnitude of refraction $r$ at a given wavelength
relative to a fixed `guide' wavelength developed by Enrico Marchetti for ESO.\footnote{See http://www.eso.org/gen-fac/pubs/astclim/lasilla/diffrefr.html}  The direction of the refraction is along the local altitude vector for a given star; this corresponds to the parallactic angle $\eta$
defined by the spherical triangle with vertices at the star, the celestial pole, and the local zenith.

\begin{equation}
\textrm{tan} \, \eta = \frac{\textrm{sin} \, h \, \textrm{cos} \, \phi \, \textrm{cos} \, \delta}{\textrm{sin} \, \phi-\textrm{sin}\, \delta \, \textrm{cos} \, z}
\end{equation}
where $h$ is the hour angle ($h > 0$ towards the West), $\phi$ is the local latitude ($\phi = 32^{\circ} \, 46' \, 49''$ for APO), $\delta$ is the target declination,  $z$ is the zenith distance  $\textrm{cos} \, z = \textrm{sin} \, \phi \, \textrm{sin} \, \delta + \textrm{cos} \, \phi \, \textrm{cos} \, \delta \, \textrm{cos} \, h$, and $\eta$ is defined in the range $-180^{\circ}$ to $+180^{\circ}$.

The SDSS 2.5-m telescope is equipped with an alt-az mount and all plates are observed with a position angle of $0^{\circ}$, so the amount of
refraction in focal-plane coordinates\footnote{SDSS xfocal/yfocal coordinates are defined such that 
$+$xfocal corresponds to $+$right ascension and $+$yfocal corresponds to $+$declination.}
is given by
\begin{equation}
\Delta x_{\rm focal} = -r \, \textrm{sin} \, \eta
\end{equation}
\begin{equation}
\Delta y_{\rm focal} = -r \, \textrm{cos} \, \eta
\end{equation}
i.e., at transit $h=0$, $\eta = 0^{\circ}$, and hence the entirety of the apparent refraction is along the yfocal direction.\footnote{In the
present work we neglect the relatively small effect of distortions introduced by the SDSS 2.5m optical system; these are, however, accounted for
in the actual data pipeline described by Law et al. (in prep).}

\begin{figure}
\plotone{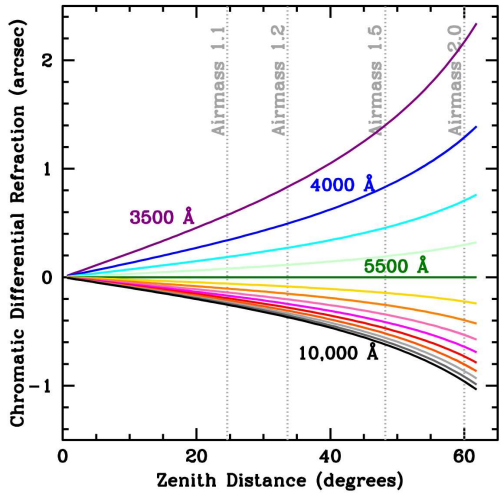}
\caption{Differential atmospheric refraction in arcsec of altitude relative to 5500 \AA\ for the MaNGA wavelength range as a function of zenith distance.  Calculations
assume median conditions for APO with air temperature 10.5$^{\circ}$ C, 24.5\% relative humidity, and atmospheric pressure of 730 mbar.}
\label{darfig.fig}
\end{figure}

Since differential refraction (particularly shortward of 4000 \AA) can be substantial
compared to the fiber radius of 1 arcsec (see Figure \ref{darfig.fig}) the spectrum recorded by a single fiber is not strictly the spectrum of a 
single region in a given galaxy; it is a
bent 'tube' that traces different regions of the galaxy at different wavelengths.  Most immediately, this means
that the effective on-sky footprint
of the MaNGA IFUs can be shifted by up to $\sim$ 1 arcsec
between blue and red wavelengths, requiring that the MaNGA
data reduction pipeline (DRP; see Law et al. in prep) rectify the spectra to a common astrometric grid when reconstructing the data cubes.
More problematically, since the three exposures in a given dither set 
will be obtained at different hour angles the relative offset at a given wavelength will change between these three exposures and degrade the intended dither pattern coverage.


\begin{figure*}
\plotone{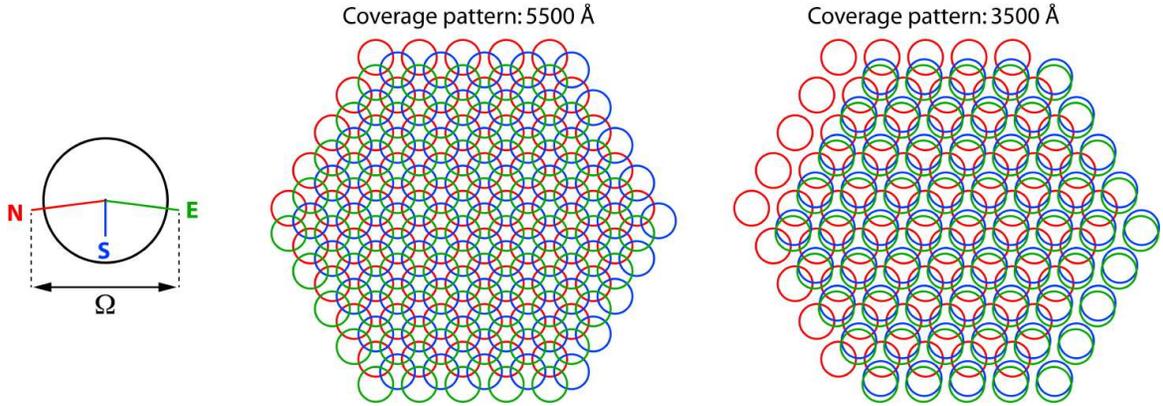}
\caption{Illustrative figure showing degradation of the intended dither pattern due to chromatic differential refraction.  In this
example we assume a  target at $\delta = +60^{\circ}$ was observed with a standard N-S-E dither pattern,
but the three exposures were taken at hour angles  $h=-4$h, 0h, and +4h respectively (corresponding to parallactic angles
$\eta = -97^{\circ}, 180^{\circ}, +97^{\circ}$).  The image on the left shows the offset due to chromatic refraction at 3500 \AA\
relative to the nominal center of a given fiber and defines the regularity statistic $\Omega$.  While the achieved dither pattern is nominal at the guide wavelength (central panel),
at 3500 \AA\ the fibers in positions S and E lie almost atop each other (right panel).}
\label{dareffect.fig}
\end{figure*}

At the guide wavelength of 5500 \AA, the three dithers will be executed properly.  As illustrated in Figure \ref{dareffect.fig} however,
at other wavelengths there will be variable shifts of the effective dithering pattern.  
These shifts can in some cases be comparable to the dither distances themselves, thereby degrading the effective dither pattern such that entire
dither postions can be effectively `lost' at certain wavelengths.  As suggested by Figures \ref{bundlesims1.fig} and \ref{bundlesims2.fig}
this produces substantial and undesirable non-uniformities in the reconstructed image depth and recovered FWHM profile across the face of each IFU.


\subsection{Field Differential Refraction}
\label{ar.field.sec}

\begin{figure}
\plotone{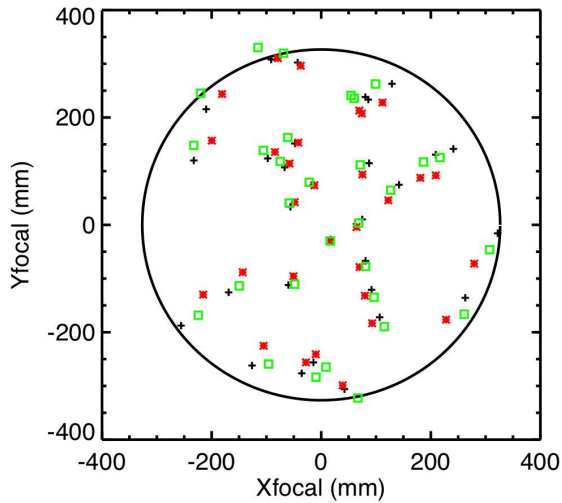}
\caption{Magnified illustration of the effects of field differential refraction at the guide wavelength ($\sim$ 5500 \AA) across the 3$^{\circ}$ diameter SDSS
plugplate.  Black `+' symbols indicate the nominal positions in focal plane coordinates of a randomly-selected set of 30 target galaxies.  These locations are computed assuming that the plate center has 
declination $+7^{\circ}$ and is observed at transit ($h = 0$ hours);
these correspond to the locations of the physically drilled holes in the plugplate into which the MaNGA IFUs are inserted.  If the same plate were observed 4 hours later ($h = +4$ hours) the apparent locations of the galaxies in the focal plane would be different due to field differential refraction both before (red asterisks) and after (green open boxes) guider corrections have been applied.
Note that all offsets from the nominal positions have been magnified by a factor of
300 to enhance visibility; the maximum actual shift after guider corrections in this example is $\sim 2$ arcsec.}
\label{allfield.fig}
\end{figure}

In addition to varying with wavelength, both the magnitude and the direction of atmospheric refraction vary according to the
location of an object on the sky, and the $3^{\circ}$ SDSS plugplate field over which our IFUs are distributed is
sufficiently large that this variation cannot be neglected.  As a given field rises, transits, and sets, the apparent locations
of astronomical targets in the telescope focal plane shift.  
As described in \S \ref{hardware.sec},
the SDSS telescope guider system compensates for this 
using guide fibers placed on astrometric standard stars distributed throughout a given field, and adjusts the overall shift, 
rotation, and scale of the  focal plane to compensate.  However, since
field compression occurs along only a single
direction (altitude) it cannot be fully corrected by a global change in the focal plane scale, leaving a residual 
quadrupole term in the guider-corrected focal plane locations of the target galaxies (see Figure \ref{allfield.fig}).\footnote{Field
differential refraction is calculated using the SDSS plate design code located on the collaboration SVN repository.}

Such field differential effects are most noticeable when observing with single fibers or an array of slits \citep[see, e.g., discussion by][for the 16' x 16' VIMOS field of view]{cuby98} since targets can rapidly shift out of the aperture.  Hence, previous generations of SDSS that have used single fiber spectroscopy have been careful to observe at hour angles close to that for which a given plate is drilled.  In contrast, MaNGA is
relatively insensitive to shifts in the effective centroid of an IFU since such shifts are small
compared to the total field of view of each IFU ($\sim 30$ arcsec for the 127-fiber IFUs).\footnote{This effect is more important for the 
spectrophotometric minibundles which have a diameter of only 7.5 arcsec; as discussed by Yan et al. (in prep), large
offsets of the spectrophotometric standard stars from the center of the calibration minibundles due to a combination of differential refraction, dither offsets, and 
other effects can complicate flux calibration.}
The MaNGA plugplates are therefore all drilled for transit ($h=0$ hours), so the holes into which the MaNGA IFUs are inserted correspond
to the expected focal plane locations of the galaxies at this point in time.

More important for MaNGA is the {\it change} in field differential refraction between exposures in a given dither set, which leads to degradation
of the effective dither pattern akin to what was seen for chromatic differential refraction in Figure \ref{dareffect.fig}.  
As illustrated by Figure \ref{allfield.fig}, the magnitude of this effect depends on the field declination, the hour angle $h$ of exposures
within a given set, and the location of an IFU within the plugplate.  In the extreme example shown in Figure \ref{allfield.fig} (low declination,
with exposures obtained many hours apart) the shift can be comparable to a fiber diameter.  In more realistic and typical cases
(field center at $\delta = +40^{\circ}$, observed at $h=0$ and $h=+1$ hours) the shift after guider corrections is typically $\lesssim 0.1$ arcsec.


\section{The Uniformity Statistic $\Omega$}
\label{omega.sec}

Given the presence of both chromatic and field differential refraction,  {\it no two exposures} taken by MaNGA will have an identical 
fiber sampling pattern even in the absence of dithering.  The primary driver of the MaNGA observing strategy is therefore mitigation of the impact of 
atmospheric differential refraction on the regularity of the dither pattern in order to achieve maximally-uniform data
quality and depth within a given IFU.

Given any two exposures separated by a time $\Delta t$ there are vectors $\vec{r_1}$ and $\vec{r_2}$ defining
the effective offset of a fiber from its intended location on the target galaxy due to chromatic differential refraction, and $\vec{s_1}$ and $\vec{s_2}$
the offset due to uncorrectable field differential refraction effects.
In our rectilinear focal-plane coordinate system the total  shifts from differential refraction are given by:
\begin{align}
\Delta x_1 = r_1 \, \textrm{sin} \, \eta_1 + \vec{x} \cdot \vec{s_1} \\
\Delta y_1 = r_1 \, \textrm{cos} \, \eta_1 + \vec{y} \cdot \vec{s_1}  \\
\Delta x_2 = r_2 \, \textrm{sin} \, \eta_2  + \vec{x} \cdot \vec{s_2} \\
\Delta y_2 = r_2 \, \textrm{cos} \, \eta_2 + \vec{y} \cdot \vec{s_2} 
\end{align}
where $\eta_1$ and $\eta_2$ are the respective parallactic angles for the two exposures
and the vectors $\vec{s_1}$ and $\vec{s_2}$ are each projected into their components along the x/y focal plane coordinate system.
The quantity of interest for survey planning purposes is the total distance between these shifted locations in the focal plane:

\begin{equation}
\Omega = \sqrt{(\Delta x_1 - \Delta x_2)^2 + (\Delta y_1 - \Delta y_2)^2}
\end{equation}

In practice, we calculate $\Omega$ between the first and last exposures in a dithered set of three frames (see illustrative diagram in Figure \ref{dareffect.fig}).\footnote{Each
exposure is 15 minutes in length; we adopt the midpoint of each exposure as the characteristic instant for purposes of calculating $\Omega$ (although see \S \ref{omegaexp.sec}).}
Using the tools developed in \S \ref{simulations.sec} we simulate four test cases where $\Omega$ ranges from 0 to 1''.  We use the as-built MaNGA 127-fiber IFU ma024,
and assume a standard three-point (N-S-E) dithering strategy in which exposure N is shifted by $\Omega/2$ in the -Xfocal direction, 
exposure S is shifted by $\Omega/3$ in the -Yfocal direction
and exposure E is shifted by $\Omega/2$ in the +Xfocal direction (see, e.g., Figure \ref{dareffect.fig}).  Note that we are free to assume such symmetry because any shift common to all
three exposures will simply result in a translation of the entire pattern.

\begin{figure*}
\plotone{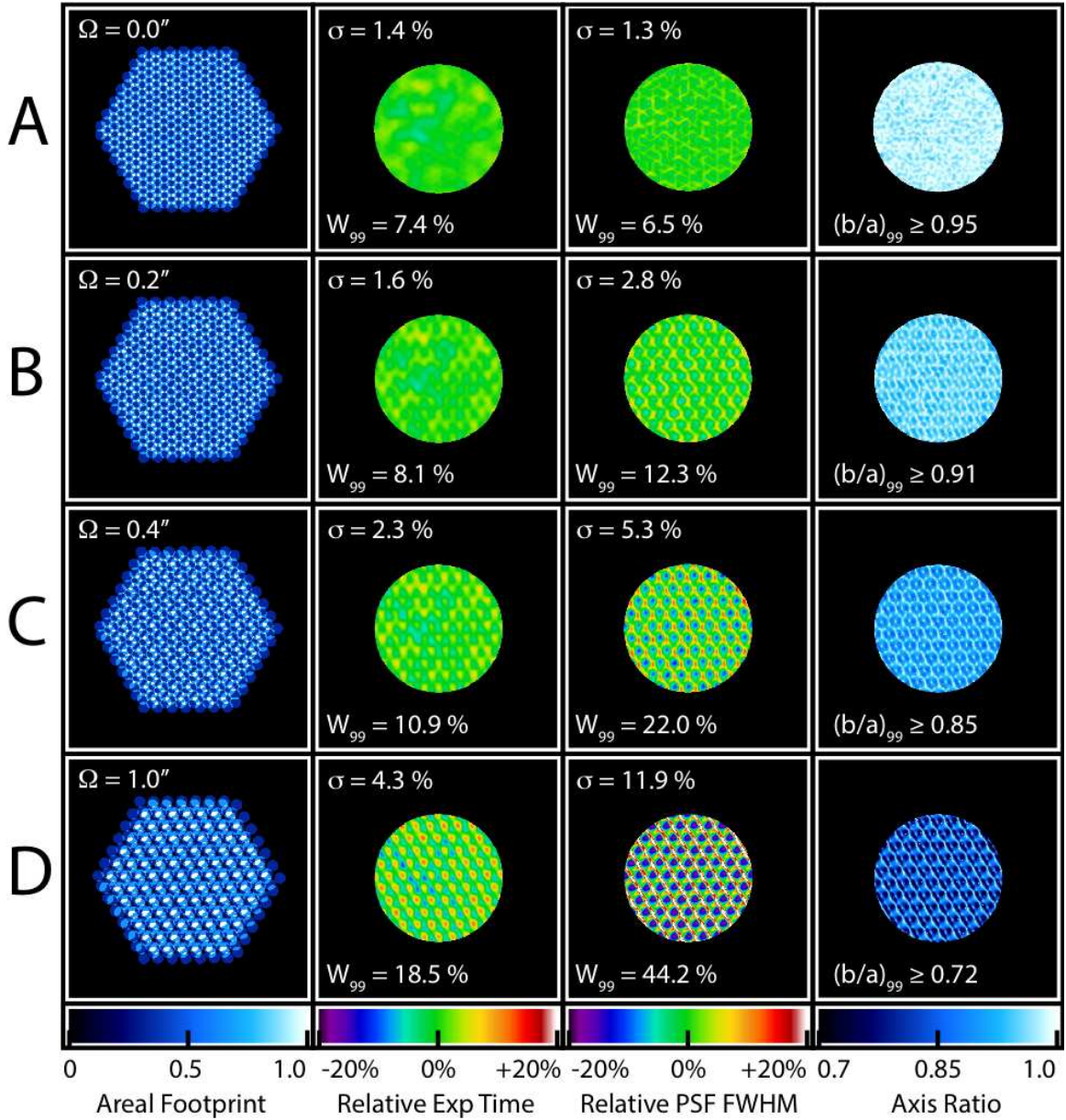}
\caption{As Figure \ref{bundlesims1.fig}, but showing simulated point-source response
variability as a function of location in an IFU for dithered observations of MaNGA 127-fiber bundle
ma024 with different values of the pattern degradation $\Omega$.}
\label{omegaeffect.fig}
\end{figure*}

We show results for the expected exposure time, reconstructed PSF FWHM, and reconstructed axis ratio uniformity as a function of $\Omega$ in Figure \ref{omegaeffect.fig}.\footnote{Note that
while $\Omega$ degrades the expect coverage pattern, we assume that the magnitude and direction of all of these shifts are known 
(see discussion by Law et al. in prep) and the true effective locations of each fiber
are used when reconstructing the data cube.}
Figure \ref{omegasims.fig} suggests that so long as $\Omega \lesssim 0.2$ arcsec observations should meet the target uniformity criteria outlined in \S \ref{regularity.sec}
with FWHM $2.65 \pm 0.08$ arcsec.
At $\Omega = 0.4$ arcsec, degradations in the reconstructed PSF uniformity and circularity start to become apparent; although the mean reconstructed PSF in the
bundle has FWHM $2.65 \pm 0.14$ arcsec the total spread of FWHM values can be as large as $\sim 0.3$ arcsec, and 99\% of locations have minor/major axis ratio
greater than 0.85.  By $\Omega=1.0$ arcsec the dither pattern is badly degraded, with reconstructed FWHM values varying by over an arcsecond depending on where
a point source falls within the bundle.
Our science requirements (\S \ref{reqperformance.sec}) therefore translate to a requirement 
that  $\Omega < 0.4$ arcsec, with the goal of reaching $\Omega < 0.2$ arcsec for the majority of observations so that it does 
not dominate the flux calibration accuracy budget.


\section{MaNGA Observing Strategy}
\label{strategy.sec}

\subsection{Set Lengths and Visibility Windows}
\label{harestrict.sec}

As described above, $\Omega$ is a complicated function of wavelength, integration time, target declination, hour angle, and location of an IFU on a given plate.
However, it is possible to define a series of relatively simple observing guidelines that will ensure that $\Omega$ stays below our 0.4 arcsec threshold.

First, we note that $\Omega$ behaves nearly linearly with the amount of elapsed time between exposures in a given set, 
meaning that  it is  desirable to obtain all three exposures in the set as close in time
to each other as possible.  Since each exposure is 15 minutes long, we therefore require that all three exposures be obtained in a
{\bf set length} of one hour  (i.e., the change in hour angle between the start of the first exposure and the end of the last exposure 
should be 1 hour or less, corresponding to 45 minutes between the effective midpoint of the first and last exposure).  
While we expect that each set of three exposures will typically last 48 minutes accounting for typical readout times
and overheads, this hour-long block provides necessary flexibility in scheduling, especially during variable weather conditions.

\begin{figure*}
\plotone{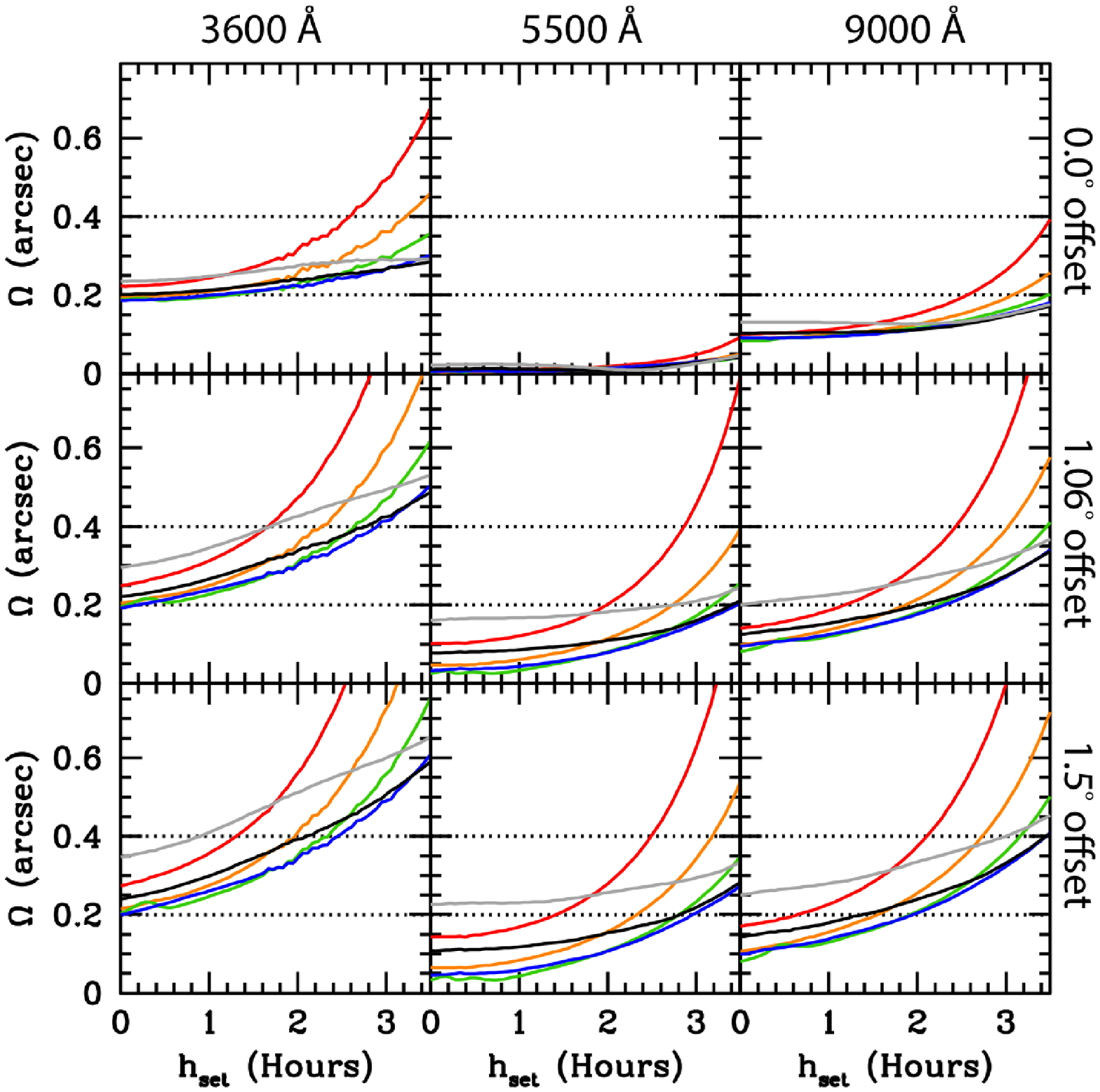}
\caption{$\Omega$ statistic as a function of midpoint hour angle of the set ($h_{\rm set}$) for a range of wavelengths, target declinations, and 
locations on a plate.  Left, middle, and right columns respectively
show results for wavelengths of 3600, 5500, and 9000 \AA; top, middle, and bottom row respectively show results for an IFU in the middle of the plate, 
$1.06^{\circ}$ towards the E edge of the plate (a circle at this radius encloses 50\% of the plate area),
and on the E edge of the plate.  Red, orange, green, blue, black, and grey solid lines respectively indicate results for 
declinations $\delta = 0^{\circ}$, $15^{\circ}$, $30^{\circ}$, $45^{\circ}$, $60^{\circ}$,
and $75^{\circ}$.  The horizontal dotted lines at $\Omega = 0.2$ and 0.4 indicate the thresholds of ideal and acceptable performance respectively.  High-frequency
structure in some lines is due to discrete changes in the best-fit guider corrections between individual simulation points.}
\label{omegasims.fig}
\end{figure*}

We next calculate the expected $\Omega$ within a 1-hour long set as a function of the midpoint hour angle $h_{\rm set}$ of the set
($h_{\rm set}$ denotes the absolute value of the hour angle midway between the start of the first and end of the last exposure).  In Figure \ref{omegasims.fig} we show the results of 
this calculation for three different wavelengths, three locations on a plate, and a range of different declinations.\footnote{Due to symmetries inherent in this exercise
(chromatic and field differential effects combining constructively or destructively), at fixed wavelength one side
of the plate will exhibit the worst $\Omega$ at positive hour angles (west of meridian) and the other at negative hour angles (east of meridian).  
For convenience
we collapse the problem such that  $h_{\rm set}$ refers to the absolute value of the hour angle, and $\Omega$ is taken to be the greater of the value
from $\pm h_{\rm set}$.}
As expected, $\Omega$ is largest  at extremely blue wavelengths
(for which chromatic differential refraction is greatest) and on the edges of a plate
(where uncorrected field differential refraction is greatest).  More importantly however,
we note that $\Omega$ grows rapidly with increasing hour angle (either East or West of the meridian) meaning that we want to obtain our observations as
close to transit as possible.
Our $\Omega$ limit therefore equates to defining a series of {\bf visibility windows} around
transit within which all MaNGA observations must be taken.

\begin{figure}
\plotone{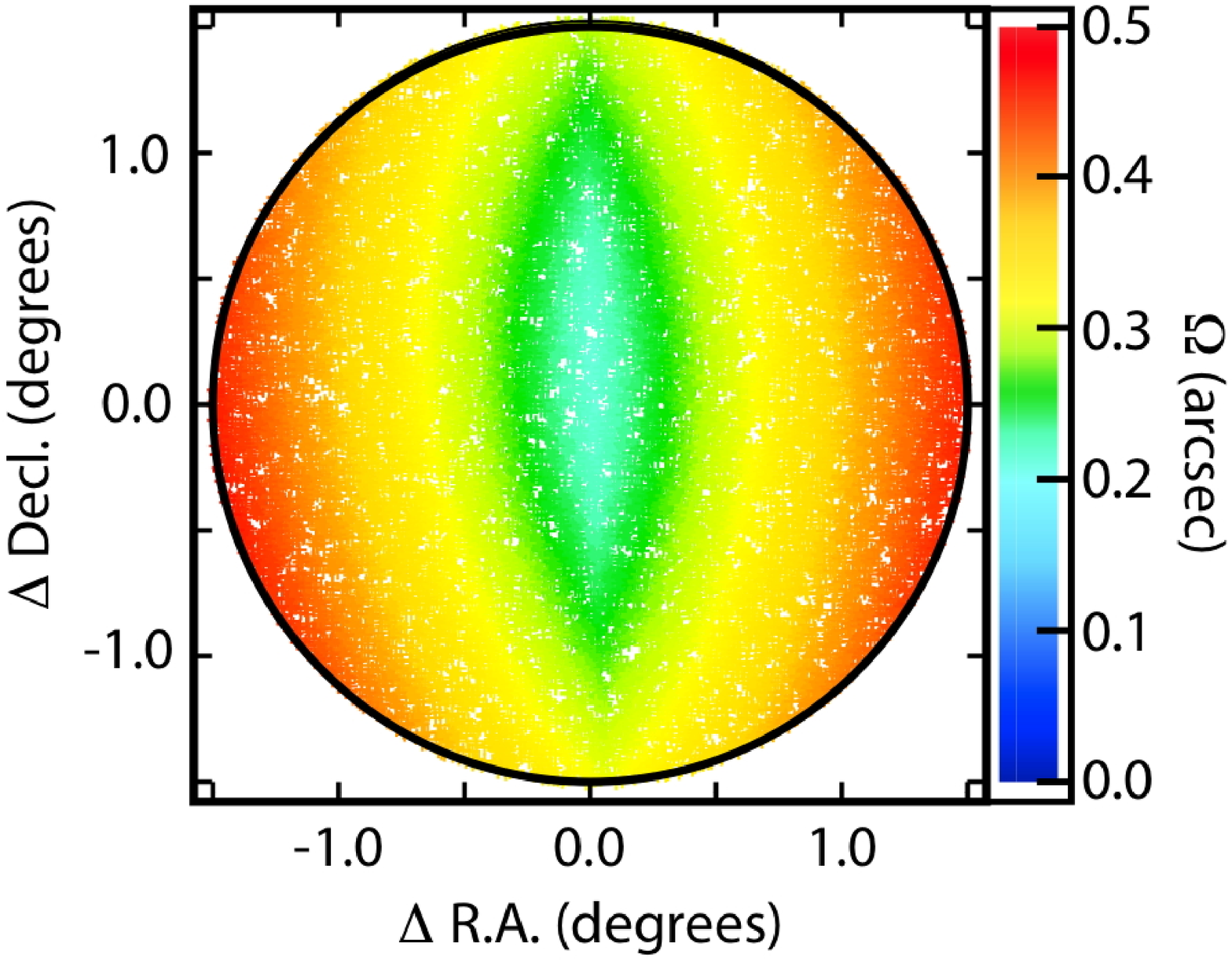}
\caption{$\Omega$ as a function of location on a plate centered at $\delta = 40^{\circ}$.  
Simulations are performed at 3600 \AA\ and assume a 5-hour observing window (i.e., $h_{\rm set} = 2.5$ hours either side of transit).
Each point represents the maximum value of $\Omega$ experienced at a given location for a hour-long set of exposures 
taken within this observing window (for one side of the plate
this maximum will occur prior to transit, for the other side it will occur after transit).}
\label{plateomega.fig}
\end{figure}

In order to compute the length of these visibility windows we require that $\Omega$ must be less than 0.4 arcsec for 
all sets, at all wavelengths, at all locations on a given plate, and at all declinations.   As indicated by Figure \ref{omegasims.fig}, the worst wavelength for $\Omega$
will be 3600 \AA, where the chromatic refraction is greatest.  We work out the worst location on a given plate as a function of declination by using Monte Carlo techniques
to compute  $\Omega$ for each of 20,000 randomly chosen locations
on an SDSS plugplate over the course of a 1-hour set.  As illustrated by Figure \ref{plateomega.fig}, the worst
$\Omega$ is typically for IFUs located on the Eastern/Western
edges of the plate for target declinations $\sim +30-40^{\circ}$; this pattern shifts at more northerly/southerly declinations.

\begin{figure}
\plotone{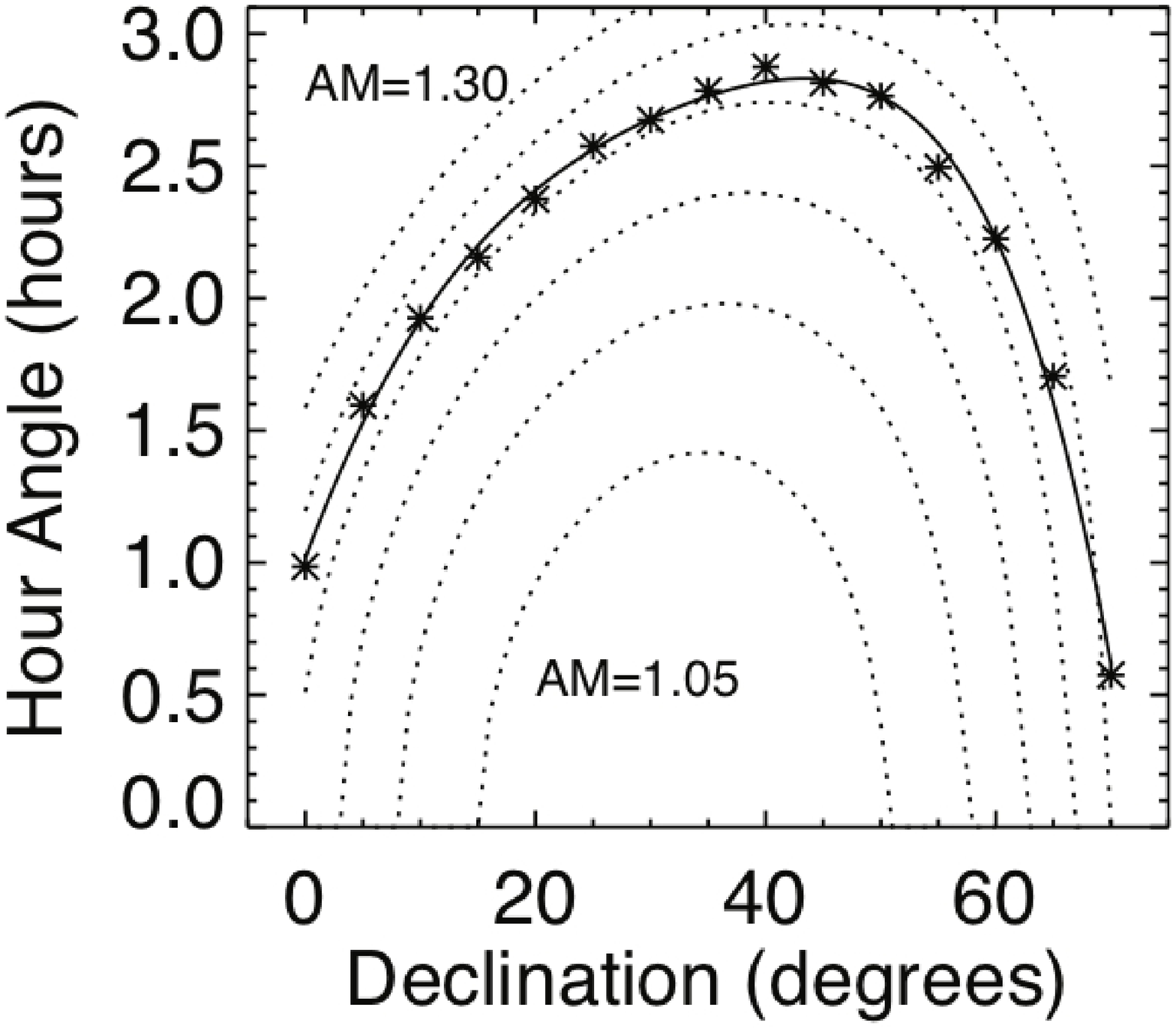}
\caption{Black asterisks show the maximum hour angle away from transit ($h_{\rm exp}$) within which all MaNGA exposures must be obtained as a function
of declination based on numerical simulations.  The solid black line represents a polynomial fit to these 15 data points.
Dotted lines indicate contours of constant airmass (every 0.05 from AM 1.05 to 1.30) 
as a function of declination and hour angle; note that these contours closely track the derived hour angle limits.}
\label{haplot.fig}
\end{figure}

Using these simulations we finally have all of the pieces required to define our visibility windows.  For a grid of declinations spaced every $5^{\circ}$
from $\delta = 0^{\circ}$ to $70^{\circ}$ we compute the limiting set hour angle such that $\Omega = 0.4$ arcsec at $\lambda = 3600$ \AA\ at the worst location
on a given plate.  Converting the set midpoint hour angle to the maximum midpoint hour angle of an individual exposure
($h_{\rm exp} = h_{\rm set} + 22.5$ minutes for 1-hour sets), we show the final visibility windows
as a function of declination in Figure \ref{haplot.fig}.  These windows range from about 1h either side of transit for fields near the celestial equator to
$\sim 3$ hours for declinations $\delta \sim +40^{\circ}$.

Intriguingly, despite all of the complications involved in computing these visibility windows they are nearly equivalent to simple airmass limits,
independent of field declination.  As illustrated by Figure \ref{haplot.fig}, our visibility windows can be described as a 6th order polynomial as a function
of declination, or more simply by the requirement that airmass AM $< 1.21$ for all exposures at all declinations.
This airmass limit is determined by the SDSS plate diameter, the BOSS spectrograph wavelength coverage, and the assumed length of each set.\footnote{It is therefore
possible to increase the airmass limit by reducing the set length or effective plate diameter (i.e., restricting the locations of IFUs on the plate).  For instance, a set length of
48 minutes instead of 1 hour would increase the airmass limit to 1.34, expanding the visibility windows significantly.  Such modifications to the observing strategy
set forth here will be actively  explored over the lifetime of the survey.}

We note that while these visibility windows have been established to ensure that $\Omega < 0.4$ arcsec 
at all wavelengths for all MaNGA observations, {\it typical}
performance is expected to be considerably better than this.  At most wavelengths,  most locations on a plate, and most hour angles within
the visibility window
 $\Omega$ will be 0.2 arcsec or below (see, e.g., Figs. \ref{omegasims.fig} and \ref{plateomega.fig}).
Additionally, these simulations have assumed that sets are completed in one hour (45 minutes between the midpoint of first and last exposures in a set).
Early survey observations at APO suggest efficiency such that most sets are actually observed in more like 48 minutes (33 minutes between the midpoint of first and last
exposures);  since $\Omega$ scales roughly linearly with the set length we therefore expect on-sky performance to 
typically be a factor $\sim 33$\% better
than assumed in these simulations.
Additionally, irregular coverage of an astronomical target in one set of exposures will tend to be averaged out across many such sets, resulting in more uniform
performance for the final data cube of a given source.


\subsection{Observing Conditions and Missing Exposures}
\label{weather.sec}

Thus far, all simulations have assumed that  atmospheric seeing remains constant throughout all exposures in a given set, and that small variations in
transparency can be normalized via per-exposure flux calibration (although see \S \ref{fluxaccuracy.sec}).
This assumption is often reasonable over the course of any given hour, but since
rapid changes in observing conditions occur on some nights we must formulate our observing strategy accordingly.

Consider, for instance, the pathological case where two dithered exposures have been successfully obtained in
good conditions, but the third is lost.  Whether it is never taken, or taken in extremely poor conditions (e.g., heavy cloud, seeing greater than 4 arcsec FWHM, etc.),
the combined set of exposures no longer uniformly samples the source image.
In such a situation, the missing exposure would have to be made up on another night, and obtained within a small range of allowable
hour angles such that the total set length is still less than one hour.

We therefore establish a series of additional requirements for image uniformity across exposures within a given set.
Based on simulations similar to those described in \S \ref{simulations.sec} and \S \ref{omega.sec} but allowing for variable 
seeing and transparency, we find that
\begin{itemize}
\item All exposures in a set should have seeing within 0.8 arcsec of each other.
\item All exposures in a set should have (S/N)$^2$ values within a factor of two of each other.
\item Each set of exposures should have median seeing 2.0 arcsec or below in order for the reconstructed image
to have FWHM less than 3 arcsec (ensuring uniformity 
of image quality between galaxies in the MaNGA survey).
\end{itemize}

Historical conditions at APO and experience during MaNGA commissioning suggest that atmospheric conditions are generally stable enough
that these criteria will not pose a serious limitation to survey operations.
In practice, exposures also can often be rearranged between sets to optimize observing efficiency and minimize the need for patching
of missing dither positions (see discussion by Yan et al. in prep),
and further modifications to the baseline strategy will continue to be explored throughout the survey.


\section{Additional Considerations}
\label{additional.sec}

Although differential refraction considerations are the primary factor that sets the MaNGA observing strategy, we also highlight a few additional
considerations here that will impact the MaNGA reconstructed image quality and must be accounted for in survey operations.


\subsection{Required Dithering Accuracy}
\label{offsets.sec}

Just as differential refraction effects degrade the effective dithering pattern and contribute to non-uniform sampling of an astronomical source, so too does
the dithering accuracy of the telescope.
As described above in \S \ref{harestrict.sec}, $\Omega$ from refractive sources will frequently be less than $0.1-0.2$ arcsec, and the individual telescope offsets must therefore
be good to better than $0.1$ arcsec in order to not be the limiting factor governing the image sampling regularity for the majority of observations.  Indeed, it is particularly important
to minimize the contribution of offsetting errors for cases with already-high $\Omega$ from differential refraction
as the compounded errors may easily  make the difference between an acceptably- versus unacceptably-uniform set of exposures.
Based on observations performed at APO during MaNGA commissioning,\footnote{The guider system uses 16 coherent imaging fiber bundles plugged on the plate and imaged
by a separate guider camera; by monitoring the positions of these 16 stars and comparing them to the desired positions we can  measure the dithering accuracy directly.}
the dither offset error has a median of 0.063 arcsec, and is smaller than 0.1 arcsec in 76\% of exposures.  Although the current dithering accuracy degrades to a median of 0.1
arcsec at altitudes higher than 80$^{\circ}$, work is ongoing to improve this performance (see details in Yan et al. in prep).


\subsection{Required Guiding Accuracy}
\label{omegaexp.sec}

In addition to the accuracy with which the telescope offsets are performed it is also important to consider the guiding accuracy of
the telescope (i.e., how well a given position is maintained over the course of an exposure).  Although poor guiding performance will not
degrade the coverage uniformity of a set of exposures, it will degrade the image quality of the exposures by 
contributing in quadrature to the effective astronomical seeing.  
Observations obtained during MaNGA commissioning show that the median guiding accuracy (based on variations in guide star positions across all 15-second
guider camera exposures during each 15-minute science exposure; see details in Yan et al. in prep) is 0.12 arcsec, substantially smaller than the median
SDSS 2.5m seeing of $\sim$ 1\farcs{5}
(computed across all BOSS spectroscopy in 2012).

We note that a similar effect is caused by differential refraction; just as changing refraction causes the effective location
of a fiber to move between two exposures (\S \ref{ar.field.sec}), so too does it cause the effective location of a fiber on a given astronomical target to move
{\it during} the exposure as well.
However, for observations obtained using the strategy outlined in \S \ref{strategy.sec} above
this effect is small.
Since differential motion of a fiber with respect to a fiducial position (i.e., $\Omega$) scales roughly linearly with time,
the motion in a 15-minute exposure will be $\sim$ 1/3 of that for a given set of 3 exposures.  Since we require the latter to be $< 0.4$ arcsec
even in extreme cases, the motion in a single exposure will be $\lesssim 0.1$ arcsec, which is small compared to the guider accuracy
and atmospheric seeing profile.


\subsection{Bundle Rotation}
\label{rotation.sec}

The rotational position $\theta$ of the MaNGA IFU bundles is controlled via clocking pins that ensure proper alignment of each IFU.
However, mechanical tolerances of the pinhole translate to a rotational uncertainty for each bundle at the level of $\sim 3^{\circ}$.
In order to ensure that individual sets meet the $\Omega < 0.2$ arcsec coverage regularity goal at the edges of the largest fiber bundles
($\sim 16.5$ arcsec radius) we specify  that the rotational offset $\Delta \theta$ required between any two exposures in a given set be
\begin{equation}
\Delta \theta < {\rm tan}^{-1} (0.2/16.5) = 0.7^{\circ}
\end{equation}

Generally, rotational tension in the IFU cables should ensure that $\theta$ remains relatively constant for a given plugging, and preliminary tests
indicate that $\Delta \theta \approx 0.2^{\circ}$ (see Law et al, in prep for further details).  However, changes in the routing path of each IFU cable
through the cartridge can lead to a different rotation 
(and small translational offsets) each time the plate is replugged, and we therefore require that sets be completed within a single
plugging.

\begin{figure*}
\plotone{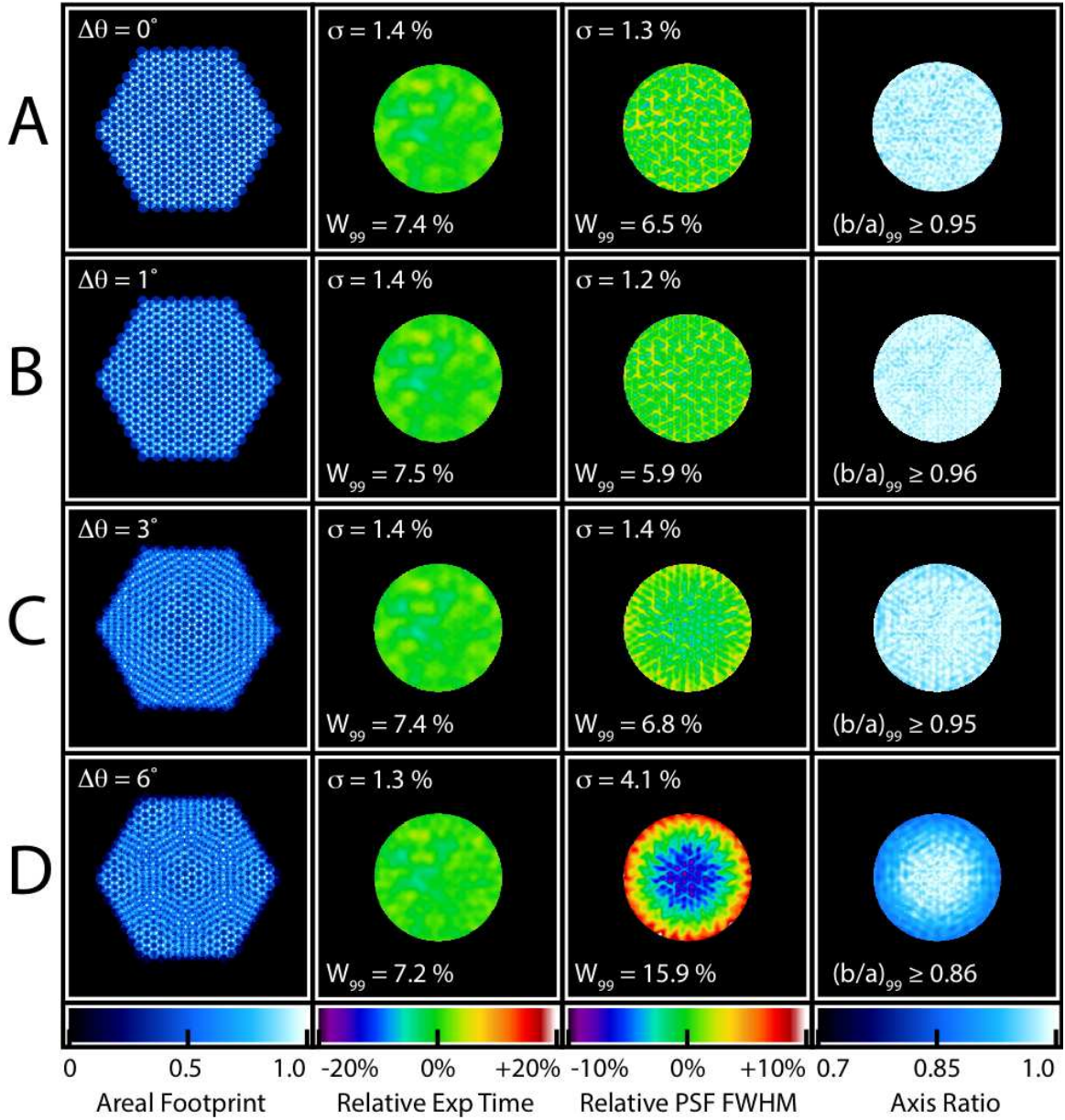}
\caption{As Figure \ref{bundlesims1.fig}, but showing the impact of uncorrected bundle rotation on the reconstructed image quality.  Note the marked increase in FWHM
and ellipticity of reconstructed point sources near the edges of the bundle for offsets $\sim 6^{\circ}$.}
\label{bundlerot.fig}
\end{figure*}

Rotation between sets of exposures can also  degrade the reconstructed image quality if it is not
measured and accounted for in the final astrometric solution.  In Figure \ref{bundlerot.fig} we simulate the effect of stacking together two sets
with different rotations without accounting for their rotational offset in the fiber astrometry.  Visible degradation of the reconstructed PSF
starts to become apparent at the edges of the IFU bundle
once the sets are rotated from each other by $\sim 3^{\circ}$, and distortions become severe once the rotation reaches $6^{\circ}$ (i.e., $\sim$ a fiber
radius at the edge of the largest bundles).
We therefore require that the MaNGA data pipeline be able to measure the rotational clocking of each IFU at the level of $\sim \pm 1^{\circ}$
so that it can be incorporated into the astrometric solution.


\subsection{Errors in Spectrophotometry}
\label{fluxaccuracy.sec}

As discussed by Law et al. (in prep) and Yan et al. (in prep), 
each MaNGA exposure is flux calibrated independently to account for variations in the atmospheric seeing and transparency.
Adequate image reconstruction is therefore dependent on the relative accuracy of the flux calibration between exposures in a given set; any offsets between exposures
will hamper the ability of the dithered exposures to properly sample the source profile.
The most pronounced effect of such offsets is not their degradation of the spatial profile of unresolved structures (e.g., point sources) however, but rather their introduction
of artificial spatial structure into a smooth background.

\begin{figure}
\plotone{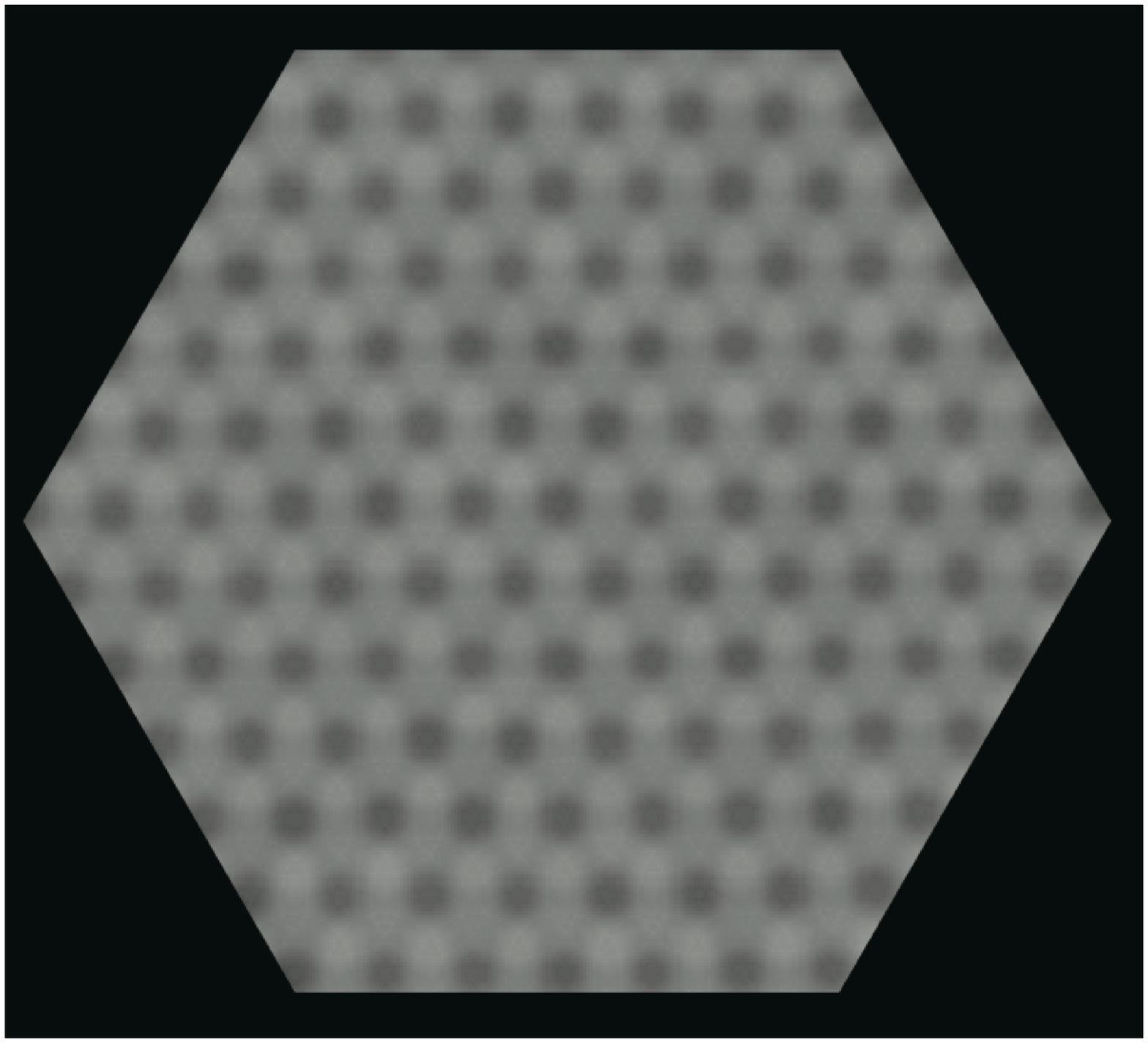}
\caption{Simulated MaNGA observation of a constant surface-brightness field showing the characteristic stippling pattern introduced by relative flux calibration
errors between dithered exposures.  Such errors introduce artificial spatial structure correlated with the dithered fiber pattern.  Greyscale stretch is arbitrary; in this
example the blackpoint (whitepoint) is set to 20\% below (above) the mean flux, corresponding to an RMS variation of about 8\% over the field of view.  Note that the simulated
field has been trimmed to omit effects from regions at the edge of the IFU field.}
\label{stippling.fig}
\end{figure}

We therefore simulate dithered observations of a constant surface-brightness source (e.g., like the outskirts of a smooth elliptical galaxy), assuming typical observing conditions
with visual seeing $\sim 1.5$ arcsec.  We mimic flux calibration errors by
multiplying the fiber fluxes for each exposure by a scale factor drawn randomly from a gaussian  distribution with a given RMS width and a median of 1.0
before reconstructing the composite image.   As illustrated in Figure \ref{stippling.fig}, calibration errors between individual exposures results
in a stippling of the smooth background, introducing artificial spatial structure correlated with the dithered fiber pattern.

In a single set of 3 exposures, we find that RMS flux calibration errors of 2\% between exposures results in a reconstructed image whose surface brightness varies
by 0.4\% (RMS) from pixel to pixel.  This is comparable to the 0.3\% pixel-to-pixel variations that we find in the reconstructed image assuming perfect flux calibration
of all exposures.  As flux calibration accuracy degrades further to 5\%, 15\%, and 50\% RMS between exposures, we find pixel-to-pixel variations of
1\%, 2\%, and 8\% respectively in the reconstructed image.  If flux calibration errors are uncorrelated between exposures in different sets, this variation will average out
over the course of observations for a given plate.  Even in the case where individual exposures are calibrated as poorly as to within a factor of 2, the pixel-to-pixel
flux for a uniform background source varies by just 2\% RMS when averaged over 4 sets (12 total exposures).
In contrast, preliminary results from MaNGA commissioning data indicate that individual exposures are typically calibrated to within 2.5\% (Yan et al. in prep), suggesting that 
flux calibration errors are unlikely to contribute significantly to the image reconstruction fidelity.


\section{Summary}
\label{summary.sec}

The MaNGA hardware design and observing strategy is driven by the desire to ensure high, uniform image quality and depth 
across all 10,000 of the galaxies that will be observed during SDSS-IV.
In particular, the goal of reaching 7\% spectrophotometric accuracy between \otwo\ $\lambda 3727$ and H$\alpha$ $\lambda 6564$
requires that the reconstructed PSF varies by 10\% or less (in both width and ellipticity) across the face of each IFU.
This goal is particularly challenging given the variable total number of exposures per field ($\sim 6-20$) required to reach the target depth, chromatic differential refraction
arising from the large wavelength coverage of the survey ($\lambda\lambda 3600-10,000$ \AA), and field differential refraction caused by the 
$3^{\circ}$ wide field of view over which individual IFUs are deployed.

We summarize the requirements necessary to meet our goal as follows:

\begin{enumerate}
\item Each IFU fiber bundle should be constructed of a regular hexagonal grid of fibers to an accuracy of $5$ \micron\ rms.
The MaNGA IFUs meet and exceed this specification with a filling factor of 56\% and a typical fiber
placement accuracy of $\sim 3$ \micron\ rms.
\item Exposures should be obtained in {\bf sets} of three 15-minute exposures dithered to the vertices of an 1\farcs{44}
equilateral triangle in order for each set to uniformly sample the image plane.
\item The telescope must be able to dither to an accuracy of 0\farcs{1} or better.
\item Each plate should be observed for an integer number of sets until the combined depth reaches a signal-to-noise ratio of 5 \AA$^{-1}$ fiber$^{-1}$ in the $r$-band continuum  at a surface brightness of 23 AB arcsec$^{-2}$.
\item All three exposures in a set must be observed within one hour of each other (i.e., the change in hour angle between the start of the first exposure and the end of the last exposure should be one hour or less), and in a single plugging of a given plate.
\item All three exposures in a set should have (S/N)$^2$ within a factor of two of each other, and be obtained in atmospheric seeing that varies by less than 0\farcs{8}.  Each set should be
obtained in median seeing of 2\farcs{0} or better.
\item All MaNGA exposures should be obtained in {\bf visibility windows} $\sim 2-6$ hours in length corresponding to airmass $\leq 1.21$.
\item MaNGA relative flux calibration between exposures must be good to $\sim 5$\% or better.
\end{enumerate}

In reality, many of the issues considered here will tend to average out over the course of the 3-4 sets that will typically be obtained on a given plate since
effects that cause a slight gap in coverage for one set will often be filled in by another.  However, our objective in designing the MaNGA program is to ensure
that the depth, coverage, and image quality of the entire survey is as good and {\bf uniform} as possible for the entire wavelength range of each of our 10,000 galaxies.
In forthcoming contributions (Law et al. in prep, Yan et al. in prep) we will explore in greater detail how well we succeed in meeting these goals.

\acknowledgements 

We thank Maryna Tsybulska for early contributions to this project, and acknowledge the Summer Undergraduate Research Program at the Dunlap Institute, University of Toronto for their support.
AW acknowledges support of a Leverhulme Trust Early Career Fellowship.
DRL thanks the anonymous referee whose comments led to an improved version of the manuscript.
Funding for the Sloan Digital Sky Survey IV has been provided by the
Alfred P. Sloan Foundation and the Participating Institutions. SDSS-IV
acknowledges support and resources from the Center for
High-Performance Computing at the University of Utah. The SDSS web
site is www.sdss.org.

SDSS-IV is managed by the Astrophysical Research Consortium for the
Participating Institutions of the SDSS Collaboration including the
Carnegie Institution for Science, Carnegie Mellon University, the
Chilean Participation Group, Harvard-Smithsonian Center for
Astrophysics, Instituto de Astrof\'isica de Canarias, The Johns Hopkins
University, Kavli Institute for the Physics and Mathematics of the
Universe (IPMU) / University of Tokyo, Lawrence Berkeley National
Laboratory, Leibniz Institut f\"ur Astrophysik Potsdam (AIP),
Max-Planck-Institut f\"ur Astrophysik (MPA Garching),
Max-Planck-Institut f\"ur Extraterrestrische Physik (MPE),
Max-Planck-Institut f\"ur Astronomie (MPIA Heidelberg), National
Astronomical Observatory of China, New Mexico State University, New
York University, The Ohio State University, Pennsylvania State
University, Shanghai Astronomical Observatory, United Kingdom
Participation Group, Universidad Nacional Aut\'onoma de M\'exico,
University of Arizona, University of Colorado Boulder, University of
Portsmouth, University of Utah, University of Washington, University
of Wisconsin, Vanderbilt University, and Yale University.

\end{document}